\documentclass[manuscript,screen,noacm,acmsmall]{acmart}
\AtBeginDocument{%
  \providecommand\BibTeX{{%
    \normalfont B\kern-0.5em{\scshape i\kern-0.25em b}\kern-0.8em\TeX}}}

\setcopyright{acmcopyright}
\copyrightyear{2024}
\acmYear{2024}
\acmDOI{XXXXXXX.XXXXXXX}

\begin{document}

{\title[Who Puts the ``Social'' in ``Social Computing''?]{Who Puts the ``Social'' in ``Social Computing''?: Using A Neurodiversity Framing to Review Social Computing Research}

\author{Philip Baillargeon}
\email{pabaill@stanford.edu}
\affiliation{%
  \institution{Stanford University}
  \city{Stanford}
  \state{California}
  \country{USA}
  \postcode{94305}
}
\author{Jina Yoon}
\email{jyoon963@uw.edu}
\affiliation{%
  \institution{University of Washington}
  \city{Seattle}
  \state{Washington}
  \country{USA}
  \postcode{98195}
}
\author{Amy Zhang}
\email{axz@cs.uw.edu}
\affiliation{%
  \institution{University of Washington}
  \city{Seattle}
  \state{Washington}
  \country{USA}
  \postcode{98195}
}

\renewcommand{\shortauthors}{Baillargeon, Yoon, and Zhang}

\begin{abstract}
Human-Computer Interaction (HCI) and Computer Supported Collaborative Work (CSCW) have a longstanding tradition of interrogating the values that underlie systems in order to create novel and accessible experiences. In this work, we use a neurodiversity framing to examine how people with ways of thinking, speaking, and being that differ from normative assumptions are perceived by researchers seeking to study and design social computing systems for neurodivergent people. From a critical analysis of 84 publications systematically gathered across a decade of social computing research, we determine that research into social computing with neurodiverse participants is largely medicalized, adheres to historical stereotypes of neurodivergent children and their families, and is insensitive to the wide spectrum of neurodivergent people that are potential users of social technologies. When social computing systems designed for neurodivergent people rely upon a conception of disability that restricts expression for the sake of preserving existing norms surrounding social experience, the result is often simplistic and restrictive systems that prevent users from ``being social'' in a way that feels natural and enjoyable. We argue that a neurodiversity perspective informed by critical disability theory allows us to engage with alternative forms of sociality as meaningful and desirable rather than a deficit to be compensated for. We conclude by identifying opportunities for researchers to collaborate with neurodivergent users and their communities, including the creation of spectrum-conscious social systems and the embedding of double empathy into systems for more equitable design.
\end{abstract}




\maketitle

\section{Introduction}

Approximately ten years ago, several landmark papers in HCI encouraged the application of neurodiversity and critical disability studies as framings for conducting research with neurodivergent (ND) users towards more accessible technologies \cite{mankoff_disability_2010, rogers_does_2013, dalton_neurodiversity_2013}. 
Research with neurodivergent\footnote{Our paper adopts the convention that "neurodiverse" refers to a group of people within the neurodiversity paradigm and "neurodivergent" refers to an individual. Some previous work has stated that "neurodiverse" should be reserved for groups of neurodivergent and neurotypical people \cite{das_towards_2021}. We adopt the view that any group with a neurodivergent person demonstrates some neurodiversity and is thus neurodiverse.} participants has historically been conducted with a deficit-based, medical lens that seeks to eliminate non-normative behaviors \cite{mankoff_disability_2010, shew_ableism_2020}. In contrast, the framework of neurodiversity states that cognitive difference, rather than being a condition that should be rehabilitated, is a natural expression of divergent human evolution and should be respected in the same manner as other protected classes (e.g., race, ethnicity, sexual orientation) \cite{med_what_2021, singer_neurodiversity_2017}. 

A neurodiversity frame that respects diverse forms of cognition and expression is particularly important when it comes to designing accessible social technologies, as networked, technology-mediated interdependence is a fixture of daily life for disabled\footnote{In accordance with worldwide language preferences of neurodivergent people, the authors primarily use identity-first language (e.g., ``autistic person'') rather than person-first language (e.g., ``person with autism'') unless the person being referred to expresses a preference for person-first language. Identity-first language is preferred by most neurodivergent people because it is conscious of the fact that neurodivergence is inseparable from the person mentioned \cite{sharif_should_2022}. Although the authors adhere to the recommendations of established language guides, we support disabled people and their right to be referred to with language that they choose and best represents their experience.} and neurodivergent people \cite{bennett_interdependence_2018, williams_cyborg_2023}. Neurodivergent people often rely on social technologies for a wider variety of tasks than neurotypical (non-neurodivergent) people to meet their needs, such as requesting essential accommodations, working remotely due to access barriers, and expressing themselves authentically in non-traditional ways \cite{williams_cyborg_2023, das_towards_2021, barros_pena_my_2023}. 
They also often have goals and values for social fulfillment that deviate from what is implied by the design of social computing systems, such as a preference for interest-based over individual-based sociality, disdain for phatic ``small talk'', and specific sensory needs that in-person social experiences fail to accommodate \cite{barros_pena_my_2023, syharat_burnout_2023, page_perceiving_2022}. These differences shape how people use social computing systems, and prior work in HCI and CSCW has developed systems that can facilitate social experience between people with these diverse values and needs \cite{barros_pena_my_2023, shin_talkingboogie_2020, stefanidi_children_2023}.

In this review, we assess whether calls for reform in the HCI community towards embracing neurodiversity have been successful in the area of social computing by examining a decade of research related to social computing systems for neurodivergent populations. 
While there have been investigations of HCI research related to autism and social computing \cite{williams_i_2021}, ADHD and technology \cite{spiel_adhd_2022}, and general accessibility \cite{mack_what_2021}, there has not yet been a review of social computing for neurodivergent users at large. 
Our review questions the values for social experience encoded by social technologies and whether implied neurotypical social values have a bearing on the methods used to construct and analyze research questions.
In other words, we explore both the systems which allow us to be social and assumptions of “being social successfully” used by researchers that design and evaluate these systems.

To understand to what degree social computing research engages with neurodiversity, we characterize the HCI community's current conceptualization of the imagined neurodivergent user of social computing systems, the problematic assumptions embedded in this imagination, and the ways in which methods of research (e.g., recruitment, level of participation of neurodivergent people) influence the generated artifacts and analysis of social experience. Through a systematic literature review of 84 papers on social computing systems designed for or used by neurodiverse groups of people, we found that the research is largely medicalized, adheres to historical stereotypes of neurodivergent children and their families, and is insensitive to the wide spectrum of neurodivergent people that are potential users of social computing technologies. We also assess the corpus by categories of research areas (e.g., systems for social networking, augmentative and alternative communication, digital collaboration) to present more detailed insights and implications. 

These findings and others \cite{spiel_adhd_2022, spiel_purpose_2021, williams_i_2021, williams_counterventions_2023} indicate that HCI research of the last several years still adopts a deficit-based model of disability that seeks to substitute neurodivergent desires with behaviors that are consistent with neurotypical conceptions of ``being social'', showing that the calls for reform ten years ago have not been wholly adopted.
We argue that truly embracing a view of social computing that respects and celebrates neurodiversity requires and benefits from genuine collaboration with existing communities. Some of these communities have existed for decades and have already built infrastructures to satisfy the social needs of their communities \cite{silberman_neurotribes_2016} that could inform social computing research. We conclude with recommendations for a research agenda that promotes neurodiversity by proposing methodological reforms consistent with critical disability theory and identifying possible features of neurodiverse social computing systems that warrant future work with neurodiverse collaborators.

\section{Background and Related Work}
To begin, we provide historical context for neurodiversity as an intersectional movement by self-advocates with aspirations of fulfillment and self-efficacy, as well as its prevalence in HCI and social computing specifically. With this context, we can then engage with related work as to develop the research questions we wish to address through our review.

\subsection{Neurodiversity and Disability Studies in HCI}
Although the term neurodiversity has its origins in an undergraduate thesis by autism advocate Judy Singer in 1998 \cite{singer_neurodiversity_2017}, the spirit of the movement has existed long before the conception of this term. Five years earlier, in 1993, Jim Sinclair published an essay “Don’t Mourn For Us” in the Autism Network International newsletter \cite{pripas-kapit_historicizing_2020}. Sinclair’s writings represented a radical break from depictions of autism as a shell or affliction that entrapped an infantile human (popularized by Bettelheim’s “Empty Fortress” \cite{bettelheim_empty_1967}), instead encouraging parents to see the children in front of them and accept them for who they are \cite{pripas-kapit_historicizing_2020}.

Research with neurodivergent participants in HCI has existed for decades, but the neurodiversity framing was not formally introduced until the early 2010s. This process began with the popularization of critical disability studies through Mankoff et al.’s contribution to ASSETS 2010, in which the authors call for a methodological shift from “mainstreaming” neurodiverse people to addressing their needs as they are \cite{mankoff_disability_2010}. Dalton continues to advocate for this methodological shift in 2013 \cite{dalton_neurodiversity_2013}, encouraging a neurodiversity framing for designing social systems. In this same year, Rogers and Marsden encouraged researchers to use a “rhetoric of engagement” rather than a “rhetoric of compassion” to position themselves as collaborators rather than ``technosaviors'' \cite{rogers_does_2013}. These publications represent a shift from a medical model of disability to a social model, which considers disability and in-access to be a socially constructed mismatch between disabled needs and infrastructure.
The choice of 2013 as the beginning of our review period was informed by the publication time period of these articles.
 
 Present usage of neurodiversity framing in HCI ranges from Dalton’s assertion of the superiority of the neurodivergent condition to a neutral shorthand for several diagnostic categories. Whereas the neurodiversity of Singer and Blume \cite{blume_neurodiversity_1998} in the late 20th century was rooted in ecological processes like biodiversity and heavily autism-centric with a positive view of autistic experience, present usage of the term often focuses on the social dimension of having needs and desires beyond the norm in a value-neutral manner. However, usage of the term varies greatly depending on the context. Our project will operate on a definition by Baumer and Freuh which defines neurodiversity as: “the idea that people experience and interact with the world around them in many different ways; there is no one ‘right’ way of thinking, learning, and behaving, and differences are not viewed as deficits” \cite{med_what_2021}. While our corpus consists of \textit{neurodiverse} social computing (ND social computing) systems, or systems that have users with a variety of neurotypes, we find that few of these systems embed the value of \textit{neurodiversity}, or the equitable involvement of users with these different neurotypes that does not privilege neurotypical values.
 
 Similar to how use of the term neurodiversity can have a variety of implications, the question of what constitutes a neurodivergent identity and what does not is largely undefined. While the authors focused on common disabilities that have a direct impact on expression, we acknowledge that our list is not exhaustive enough for some, and it may be too broad for others by including conditions that might be temporary (e.g., some forms of intellectual disability) or can result from injury prior to birth or early in childhood (e.g., cerebral palsy). We include temporary, acquired, and developmental conditions for two reasons. First, all conditions that surfaced in our search impacted social fulfillment and did so in similar ways. Whether a deviation from normative cognition occurs at six months or sixty years, no person’s thoughts are defective and no one who speaks, no matter the method, should be silenced. Second, in observance of ``crip time'', we believe that ability is temporal and access is deserved by all people, regardless of how they experience time \cite{hamraie_crip_2019}. The neurodivergent identities discussed in this paper can be found in the Appendix (8.2.5).
 
\subsection{Neurodivergent Sociality}
Building from Ochs and Solomon’s study of autistic sociality \cite{ochs_autistic_2010}, the concept of neurodivergent sociality suggests that ND people may have social practices not satisfied by traditional social environments. ND people have social needs and desires; while autism’s root in the Greek “autos” suggests apathy toward social interaction, many autistic people (and more generally, ND people) may prefer exchanges that differ from traditional face-to-face verbal interactions. This misalignment in needs between ND and neurotypical (NT) people is multifaceted and bidirectional. Two social theories, Muted Group Theory (MGT) and double empathy, suggest that realignment requires a holistic approach that cannot separately address the needs of NT and ND people. Muted Group Theory \cite{meares_muted_2017} states that the language of the dominant group (neurotypical conversational norms) makes it difficult for ND and NT people to coordinate socially. For example, a social event  defaulting to an in-person option can make it hard for someone with another preference to ask for alternative modes of engagement, even when they are reasonable (e.g., viewing a lecture remotely with closed captioning). Similarly, double empathy critiques the attribution of social difficulties for ND people to a lack of theory of mind, or a lack of awareness for their conversation partner \cite{baron-cohen_theory_2000}. A theory of mind framing does not acknowledge the real harm conversation partners cause to ND people when they are unwilling to adapt to and acknowledge ND speech. Double empathy provides a more holistic view by considering a lack of convergence of ND and NT social practices as the principal factor that complicates social coordination instead of the negligence of an individual \cite{milton_ontological_2012}.
 
Communication is not solely dependent on the members of a conversation; sensory factors in the environment (e.g., harsh lights, loud noises, crowded public transportation) can be the difference between access and obstruction for ND people in social settings. For some neurodivergent people, sensory differences and related motor disability can make physical social spaces practically inaccessible, 
requiring ample foresight and meticulous research \cite{kenna_neurodiversity_2023}. 
The constant necessity of “masking”, or attempting to hide neurodivergent traits, when in inaccessible settings can contribute to burnout and effectively bar neurodivergent people from certain careers irrespective of their expertise \cite{syharat_burnout_2023}.

\subsection{Recruitment in Neurodiversity Research}
The disability community has adopted the refrain “nothing about without us” to draw awareness to the gap between researcher goals and the needs of disabled people and the lack of representation and participation in research.
Unfortunately, recruitment of ND participants is often done without acknowledgement of its problematic history and produces work that is unrepresentative of the diversity of the ND community \cite{albrecht_handbook_2001}. Recruitment schemes that are restricted to specific venues are only representative of a particular subset of this community. For example, engaging with ND young adults in skill training centers and their support structures may lead to insights for government programs or technologies that facilitate learning. These facilities are also  important for the well-being of some neurodivergent people, as they provide assistance that allows them to live with some autonomy. However, participants recruited from these locations are not representative of all ND people (even those under the same diagnostic criteria). It is also well documented that support centers often choose “expert witnesses” or “prime subjects” when asked for participants for academic studies \cite{archibald_challenges_2015}. In the social computing context, this may lead to the exclusion of “less social” people, including those who are nonverbal or minimally verbal. Somewhat ironically, this could eliminate people who frequently rely on technology to communicate from studies of technology mediated communication.

\subsection{Neurodiversity and Social Computing}
Social computing systems and neurodiversity have been deeply linked since their inception. Not only did the passage of the ADA and autism movements of the 1990’s coincide with the proliferation of widespread consumer internet, but Singer’s thesis in which she proposed the term “neurodiversity” focused on a study of \textit{online} autism communities \cite{singer_neurodiversity_2017}.
The existence and success of sites across the history of the internet (from Wrong Planet to neurodivergent Reddit communities) suggest that the infrastructure of online social platforms has the potential to cater to neurodivergent sociality. Many of the issues discussed above (sensory difficulties, nonverbal communication schemes, inaccessibility of physical space, etc.) can be minimized or eliminated in digital space. One study revealed that autistic adults are far more likely to use digital platforms to socialize than their neurotypical peers \cite{ahmed_constrained_2022}. Neurodivergent presence online is so pervasive that projects studying online communities have revealed the latent power of neurodivergent users to influence platform dynamics \cite{simpson_hey_2023}. This all informs our focus on both neurodiversity generally and social computing specifically as a necessary contribution to HCI research.

More recently, Williams and Park investigated the experiences of neurodivergent users of social computing systems, identifying strategies used to complete tasks along four domains---mental, embodied, social, and environmental. These included using text to coordinate reminders with communication partners, leaving digital notes to mark the passage of time, and sharing message boards to coordinate tasks with housemates. These ``vignettes'' emphasize the importance of all people involved in the social system, not just a singular neurodivergent person, and the multifaceted use cases of these systems \cite{williams_cyborg_2023}. This informed our approach in understanding to what degree ND social computing research engages with NT participants in addition to ND participants.

\subsection{Critical Perspectives on Neurodiversity in HCI}

Ample research in HCI has been conducted with neurodivergent people and has been captured in reviews with various scopes, from accessibility generally \cite{mack_what_2021} to games for neurodiverse people \cite{spiel_purpose_2021} to participatory research with autistic children \cite{spiel_agency_2019} to technology for ADHD users investigated by ADHD researchers \cite{spiel_adhd_2022}. These studies focused on a specific neurodivergent condition, a specific technology, or a specific part of the research process, whereas our work provides insight into both social computing specifically and research with neurodivergent participants generally.
This prior work also informed the design of our study. As mentioned previously, our choice of time period was informed by the publication of critical work advocating for neurodiversity in the period of 2010--2013~\cite{mankoff_disability_2010,rogers_does_2013,dalton_neurodiversity_2013}. In addition, the diagnostic standards for several neurodivergent conditions were revised in 2013 to better assess their application across age and gender and address historical biases toward young males in the diagnostic process \cite{association_diagnostic_2022, glidden_gender_2016, miller_i_2020}. 

In addition, prior critical reviews in HCI helped us to determine areas of focus for our research questions.
Biases toward autism and children in neurodiversity research found by a prior literature review of accessibility and HCI \cite{mack_what_2021} prompted our investigation into what populations are included in ND social computing research, including historically underrepresented groups in research on neurodiversity, such as ADHD and dyslexic people, neurodivergent adults, and others.
Prior work has also found that much of the participation of ND people in user experience research was restricted to the evaluation phase, potentially as a byproduct of the prevalence of children in ND technology research~\cite{corlu_involving_2017}. This suggests that ND participants cannot meaningfully impact the requirements or  design of a system before completion,informing our research question on how populations are involved in different phases of the design and research process.
In a review of games for ND people, Spiel et al. find that a majority of games are single player and are constructed with an explicit medical purpose despite the abundance and popularity of multiplayer online games as a medium for neurodivergent social experience \cite{spiel_purpose_2021}.
 In addition, Williams and Gilbert study wearable technologies for autistic users and found that, while 90\% of technologies identified deficits associated with autism that their system would overwrite with normative behaviors, only 10\% addressed identified user needs such as emotion and sensory regulation \cite{williams_perseverations_2020}.
The prevalence of a medical frame in adjacent domains led us to question whether ND social computing research also has implicit or explicit goals of correcting perceived deficits and enforcing normative behaviors.


Although recent publications have investigated ND social computing systems developed by ND people outside of a formal research setting \cite{williams_cyborg_2023}, the authors could find no previous investigation of social computing and neurodiversity generally. The evidence above suggests that existing sociotechnical systems may address some needs of some ND people, but it is unclear how ND social computing research can better design for and with underrepresented ND people in a manner that is appropriate for their varied needs and desires. We then ask the following research questions, developed to assess the ways in which ND social computing research involves ND people and the ability of the developed systems to support ND sociality:

\begin{itemize}
    \item[RQ1] What populations are included in research about neurodiverse social computing?
    \newline \textit{Although many social computing systems are designed for teenagers and adults, research with ND participants often involves children. Autism is also much more represented than any other ND condition. We would like to know if this bias toward children and autism is prevalent in our corpus.}
    \item[RQ2] What perspectives towards disability inform existing works that study ND social computing?
    \newline \textit{A medical framing often involves deficit-based language that neglects the needs of ND people in favor of enforcing normative behaviors. We wonder what the stated goals of these non-prescriptive social systems are and how they might encode social norms.}
    \item[RQ3] How are these populations included in the requirements, design, and evaluation of social computing systems?
    \newline \textit{Prior work shows that ND people are often only involved in the evaluation phase of the design process and are thus unable dictate the requirements or meaningfully impact the development of a system.}
    \item[RQ4] What are the existing design recommendations for ND social computing, and how do research perspectives influence these results?
    \newline \textit{Medicalized language often identifies the minimizing of harm as a goal. We wonder if this comes at the expense of a complete social experience, as risk taking is a necessary part of being social online.}
\end{itemize}

\section{Review Method}
We engaged in a comprehensive literature review to critically examine the past ten years of social computing research with neurodiverse participants. Papers in our corpus include artifact, empirical, and mixed method contributions. To quantify levels of participation, we have adopted the system presented by Çorlu et al. to track involvement of neurodivergent participants in the requirements, development, and evaluation phases of the research process \cite{corlu_involving_2017}. Also inspired by more recent work by neurodivergent researchers in HCI and with a consciousness of epistemic violence \cite{ymous_i_2020}, we understand that the recruitment and presentation of neurodivergent participants in research is a significant indicator of the goals with which this research is undertaken. Thus, we are interested not only in the way these technologies explicitly position themselves as helpful to neurodivergent people, but the multiple identities of the neurodivergent people included in these studies and researchers' perspectives toward their disability. By developing a complete understanding of who participated in these studies and researchers' conception of how social computing technology can contribute to neurodivergent social fulfillment, we can assess the impact of these methodological choices over time.

\subsection{Positionality}
In recognition of the practices of feminist and critical disability scholars, the authors of this work make clear the perspective with which we undertake this review. At least one of the authors of this paper identifies as neurodivergent. All authors are based in the United States and are members of western higher education institutions as students, researchers, and/or faculty. Even as we saw friends, family, and ourselves depicted in this research in demeaning ways that denied neurodivergent people agency over their bodies and behaviors, we applied our criteria as objectively and consistently as we could. Although what we found often upset and enraged us, we conducted this review because we believe that in unifying disparate attempts we can do better as researchers in the HCI community. We critique work before us not as a critique of specific authors but to interrogate the values of the community that creates a permission structure for epistemic violence \cite{ymous_i_2020}. We also understand that the disclosure of a neurodivergent identity can have negative consequences and are not suggesting that neurodiverse researchers should “out” themselves in each publication for the sake of transparency. While corporations and conferences celebrate the idea of “neurodiversity” publicly \cite{keating_neurodiversity_2022}, disclosure does not always grant material benefits to the disclosing person and thus should not be universally required.

We would like to conclude this disclosure with an acknowledgement that, whether through our inexperience with research or limited experience with neurodivergent scholarship, we may make assessments or claims that upset readers. We also believe that there are many well-intentioned researchers who make similar, well-intentioned judgments, and we hope to make our inexperience an asset that promotes discussion and necessary allyship with neurodivergent people in the HCI community \cite{liang_embracing_2021}.

\subsection{Query and Inclusion Methodology}
We query the \href{https://dl.acm.org}{ACM Full-Text Library} in the period between January 2013 and June 2023 to ascertain the state of social computing systems research.
To engage with ND sociality, we found it necessary to focus on social computing systems that provide open-ended, non-prescriptive augmentations or transmissions of speech rather than therapy application systems.

We developed the following criteria to limit our analysis to the following types of projects and systems that study neurodivergent sociality:

\begin{itemize}
    \item Survey or novel technology should be an application of social computing, or “the use of technology in networked communication systems by communities of people for one or more goals” \cite{ucsb_social_computing_group_what_nodate}
    \item Elements of the corpus are published papers (full, work in progress, poster, workshop, etc.) and published between January 2013 and June 2023
    \item Paper is focused on neurodivergent users and their social needs being communicated to a conversation partner in an open-ended, non-prescriptive way
    \begin{itemize}
        \item This excludes projects focusing on aging, skills training, and administration of therapy regimens. We wished to understand how researchers went about compiling their requirements, building a novel system, or becoming familiar with an existing one, while engaging with ND experiences with the system. Projects with the foci listed above often excluded some or all of these elements.
    \end{itemize}
    \item Technology was evaluated by some stakeholder (e.g., ND people, parents of ND children)
    \item If there are multiple publications that use data from the same sample and the same technology from the same authors, the most recent publication was selected
\end{itemize}
The query listed in the Appendix (Section 8.1) was used to generate 863 initial results. To develop our query, we independently searched for papers that fit our inclusion criteria, then developed a search query that would include all of these papers in addition to papers not found in our initial search (taking inspiration from Spiel and Gerling \cite{spiel_purpose_2021}).

Out of the 863 initial items, 488 were deemed irrelevant or insufficient from their title or abstract alone. Reasons for rejecting these papers included single user systems for language learning \cite{wadhwa_collaborative_2013}, purely diagnostic systems \cite{shelke_autism_2022}, and papers whose focus was a non-neurodivergent condition \cite{de_choudhury_anorexia_2015}. Still others were excluded because they used autism as a metaphor \cite{kaminka_curing_2013} or included acronyms with multiple meanings (e.g., ASD as “agile software development” instead of “autism spectrum disorder”) \cite{behutiye_documentation_2020}. The first author then searched through the 375 remaining articles, reading their abstract and introduction. Of the 375 articles, 124 were not related to communication or social behavior after inspection of the full article, 34 were duplicates of other articles captured by the search, 30 were platforms that did not directly facilitate communication (games without communication mechanisms, physical rooms, etc.), 16 were in formats that gave insufficient information (talk descriptions, poster write-ups, etc.), and three had body text in a language no authors are fluent in (two in Portuguese, one in German). The first author then read all 168 remaining articles in full. In the interest of analyzing open-ended ND sociality, a further 84 were removed for the following reasons: lack of functionality to support the communication of open-ended feelings or desires (n=58), no research method with stakeholders employed (n=15), and restriction to therapy contexts (n=11). The final selection was 84 papers.

\subsection{Coding Strategy}
To address our research questions, we coded each document for a mix of strict categories (sample size, identity studied, contribution type, positionality, etc.) and open-coded subject areas (design recommendations) in order to contribute proper understanding of both methods and results. After developing these categories, the first and second authors coded a randomly sampled subset of the corpus (10 papers) and revised codes with a large degree of disagreement (less than 80\% agreement for a given code applied to the subset). If sufficient agreement was not reached on a subset of papers, another subset of ten papers was sampled and coded by the first and second authors using the newly revised codes. Once a consistent set of codes was developed, the first author read and coded all of the papers. This flexible method, involving inductive coding and using subsets to refine these codes, takes inspiration from Boyatzis's version of thematic analysis \cite{boyatzis1998transforming}, later used in Spiel and Gerling's review of games research in HCI for neurodivergent players \cite{spiel_purpose_2021}. Note that, in addition to coding each of the 84 papers for overall characteristics, each of the 181 samples within these papers were also coded for several characteristics including demographics and requirements for formal diagnosis similar to prior work by Çorlu et al. \cite{corlu_involving_2017}. For example, if a paper conducted a workshop for need-finding then a user study for evaluating the artifact that was generated, both of these samples were coded for their demographic characteristics separately. See the codebook in the Appendix to view the criteria used to classify each paper.

\subsection{Limitations}
As discussed in Section 3.1, all authors approach this review with a western and privileged perspective. Our search is limited to ACM venues, meaning it is by no means representative of the totality of ND technology research. Our corpus is restricted to English language publications due to the linguistic capabilities of the authors. Additionally, our list of neurodiverse identities used in our search does not represent all neurodiverse identities. Some notable omissions include OCD, Tourette syndrome, or generalized anxiety disorder. These identities were not able to be included for a variety of reasons, including the technologies covered exclusively applying an existing therapy regime or insufficient analysis of social fulfillment by neurodiverse people. We hope future social computing research engages with these identities using the framework of neurodiversity.

\section{Corpus Overview}

\begin{table}
\small
    \centering
    \begin{tabular}{{p{0.35\textwidth}|p{0.6\textwidth}}}
    \toprule
       \textbf{Research Area}  & \textbf{Papers} \\ \midrule
        Social Network Services (SNS) (n=27, \textbf{32.14\%}) & \cite{hong_investigating_2013, xu_rove_2014, hong_towards_2015, hooper_building_2015, saha_demonstrating_2015, piper_technological_2016, austin_ascmeit_2016, wilson_put_2016, sousa_inclusive_2016, vezzoli_dyslexia_2017, neupane_social_2018, bayor_characterizing_2018, wu_design_2019, andreasen_digital_2019, bayor_leveraging_2019, fox_connection_2019, zhu_co-designing_2019, macmillan_are_2020, bayor_toward_2021, race_designing_2021, rocheleau_privacy_2022, page_perceiving_2022, mok_social_2022, barros_pena_my_2023, choi_love_2023, stefanidi_children_2023} \\ \hline
        Digital Collaboration (n=17, \textbf{20.24\%}) & \cite{mokashi_exploration_2013, borges_customized_2013, simm_prototyping_2014, sinha_development_2014, yaghoubzadeh_spoken_2015, corlu_mediated_2016, saggion_able_2017, koumpouros_user_2020, friedman_using_2018, zolyomi_managing_2019, hwang_exploring_2020, tang_understanding_2021, das_towards_2021, mynatt_pivoting_2022, venkatasubramanian_designing_2022, puhretmair_peer--peer_2022, silva_unpacking_2023} \\ \hline
        Augmentative and Alternative Communication (AAC) (n=17, \textbf{20.24\%}) & \cite{sampath_assistive_2013, chompoobutr_using_2013, santos_computational_2014, takano_affective_2014, aihara_individuality-preserving_2015, alessandrini_designing_2016, chen_effectiveness_2017, bircanin_challenges_2019, shin_talkingboogie_2020, valencia_conversational_2020, samonte_tap--talk_2020, narain_personalized_2020, robertson_designing_2021, dai_designing_2022, fontana_de_vargas_aac_2022, albert_aid_2022, scougal_perceived_2023} \\ \hline
        Sensors for Social Feedback (n=8, \textbf{9.52\%}) & \cite{pervaiz_speechometer_2014, boyd_saywat_2016, halperin_social_2016, boyd_procom_2017, crowell_variations_2018, lin_empathics_2020, chen_cognitive_2021, race_understanding_2022} \\ \hline
        Workplace Culture and Organizations (n=8, \textbf{9.52\%}) & \cite{morris_understanding_2015, zolyomi_values_2018, sharma_socio-technical_2020, balasuriya_support_2022, moster_can_2022, tymoshchuk_digital_2022, wang_invisible_2022, wang_virtual_2023} \\ \hline
        Games (n=7, \textbf{8.33\%}) & \cite{hernandez_design_2014, ringland_making_2015, ringland_would_2016, ringland_will_2016, ringland_making_2017, ringland_place_2019, k_l_nielsen_teaching_2021} \\ \bottomrule
    \end{tabular}
    \caption{Corpus elements by research area. Research areas were inductively developed to group the technologies of interest within our corpus based on common mediums of communication and stated goals. See Section 8.2.4 of the codebook in the Appendix for a description of each research area, and see Section 5 for an in-depth analysis of themes within papers of the same research area.}
    \label{corpuselems}
\end{table}

After some insight into the publication forums and cultural contexts in which these papers were constructed, we focus on the ND people involved in this research (RQ1), including who they are, where they come from, and how they are identified. We then discuss the motivations of the authors in our corpus for developing ND social computing systems for neurodivergent users and their depiction of neurodivergence (RQ2). This will allow us to determine how neurodivergent people impact research in various domains (RQ3). In the next section, we assess each of the papers by research area, as depicted in Table 1, to gain insights into trends within the design of particular social computing systems and how research methods vary between research areas (RQ4). The following categories of research areas were uncovered within our corpus, formulated based on the technology discussed and the conventions used by the subfield of HCI in which it is discussed:
\begin{itemize}
    \item \textbf{Social Network Services (SNS)} involve the ``use of Internet technologies to link together users of the Internet who have common interest'' \cite{ince_dictionary_2009}. Facebook, Reddit,  and YouTube would all be considered social network services. Papers in this category involve individual users of SNS platforms exploring their interests and for personally defined reasons.
    \item \textbf{Augmentative and Alternative Communication (AAC)} systems are most commonly used by nonverbal autistic and physically disabled people to verbally communicate by translating non-verbal inputs into verbal speech. Examples include predictive text-to-speech systems, Picture Exchange Communication Systems (PECS), and drawing boards.
    \item \textbf{Digital Collaboration} systems provide a technical context through which information can be shared between users collaborating on a shared desired outcome. Examples include teleconferencing and collaborative self regulation apps. Papers in this category involve users completing technology-mediated social tasks with colleagues, friends, or family.
    \item \textbf{Sensors for Social Feedback} systems are often visual aides that are designed to help neurodivergent people by translating latent social information from the ND user’s environment into digital cues. Examples include facial expression monitoring and distance measurement for conversation partners.
    \item \textbf{Workplace Culture and Organizations} consists of a combination of social computing systems used in skill training facilities, workplaces, and schools. Technologies discussed may include teleconferencing, collaborative IDEs, 3D printing, AR/VR experiences, and others. Analyses often focus on ND people learning how to use these systems.
    \item \textbf{Games} take the form of multiplayer, digital worlds that allow users to collaborate, compete, and communicate in highly structured, game-specific architectures. Examples include Minecraft and Counter-Strike. The context of playing the game as well as the explicit articulation of social rules, goals, and purposes through the design of the game world provide a unique social experience different from the mediated communication provided by SNS.
\end{itemize}

Research areas are detailed further in Section 8.2.4 in the Appendix, and each paper's research area is displayed in Table 1.

\subsection{Publication Forums and Research Areas}
\begin{figure}
    \centering
    \includegraphics[scale=0.6]{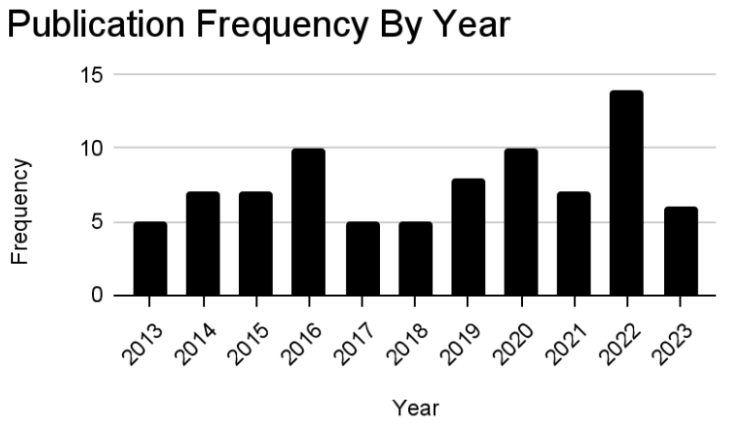}
    \caption{Publications per year for papers included in our corpus for the years 2013-2023. Note that 2023 only includes publications from January 1, 2023 to June 30, 2023.}
    \label{pubsbyyear}
    \Description[There are an average of 7.6 publications per year with no clear trend.]{The number of publications per year are: 5 in 2013, 7 in 2014, 7 in 2015, 10 in 2016, 5 in 2017, 5 in 2018, 8 in 2019, 10 in 2020, 7 in 2021, 14 in 2022, and 6 in the first half of 2023.}
\end{figure}
\textbf{ND social computing research may be growing, but evidence of this is inconclusive.} The average number of publications per year is 7.6, with a sizable increase in the number of publications since 2020, but there is no clear upward trend. An analysis of Google Trends data for the term “neurodiversity” shows a massive increase in interest worldwide beginning in late 2020, which would be consistent with the strong output in 2022 and 2023 in our corpus \cite{google_trends_neurodiversity_2023}. 
\begin{figure}
    \centering
    \includegraphics[scale=0.6]{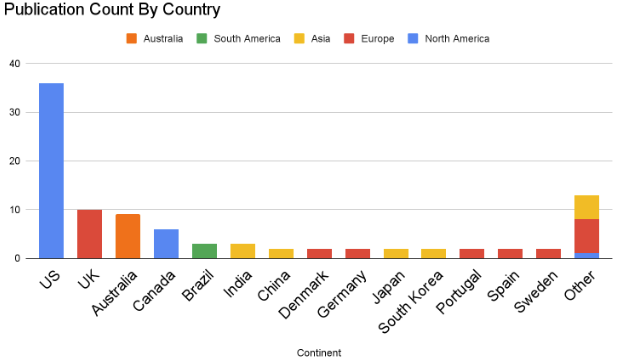}
    \caption{Publications by country, where countries with two or more publications are listed individually. Due to collaborations, bars do not sum to 84. North America, and the United States specifically, dominates discourse on ND social computing, meaning conceptions of neurodivergence that differ from those in the US, UK, and Australia are unlikely to be considered despite these systems involving users across multiple regions.}
    \label{pubsbycountry}
    \Description[Most papers are from institutions in the US, UK, Australia, and Canada.]{The number of papers with an author based in each country are: 36 from the United States, ten from the United Kingdom, nine from Australia, six from Canada, three from Brazil, three from India, two from China, two from Denmark, two from Germany, two from Japan, two from South Korea, two from Portugal, two from Spain, two from Sweden, and 13 other countries each had one publication (seven from Europe, five from Asia, one from North America).}
\end{figure}

\textbf{ND social computing papers are almost exclusively in western contexts.} Most publications originate from North American (n=49) and European (n=25) universities. Even considering the authors only analyzed English-language publications, the complete absence of the African continent in this research area is concerning. Prior work shows ND advocates in Africa and Asia are often overlooked, as self-advocacy is more rare, and cultural values differ from those that western researchers prioritize \cite{cheng_neurodiversity_2023}. Global representation in ND social computing is important because perceptions of neurodiversity can vary greatly depending on cultural context \cite{kopetz_autism_2012}, and many of the technologies in our corpus from the global south seek to combat stigma surrounding neurodivergence through communication \cite{sinha_development_2014, samonte_tap--talk_2020}. Additionally, many of the papers in non-western countries take the form of an AAC for a language spoken in this country, showing a lack of support for multiple languages and cultural contexts in the design of AAC systems.

\textbf{Conceptions of ND sociality in ND social computing are further limited by the prevalence of existing systems as the focus of papers in our corpus.} Shifting from the wider cultural context of the papers included to the types of contributions these papers provide, most contributions \cite{wobbrock_seven_2015} are empirical (n=45, 53.57\%), meaning they studied the usage of an existing technology. The next most common contribution type was artifacts (n=26, 30.95\%) in which a novel social computing system was created, followed by mixed methods (n=13, 15.48\%) which often featured an empirical need-finding phase and a constructed artifact. In terms of the types of systems produced and studied, the largest single category is unsurprisingly SNS (n=27, 32.14\%), followed by Digital Collaboration (n=17, 20.24\%) and AAC (n=17, 20.24\%). By limiting the study of ND social computing to ND experiences on existing platforms, observations then focus on experiences of ND people within the norms and rules of these systems rather than allowing for the development of novel organizing structures. In doing so, researchers risk a narrative in which ND people ``fail'' at using social computing systems rather than social computing systems inadequately addressing the needs of ND people. 
This is often the case for AACs, as there are three times as many papers that produce AACs than those that study the usage of them. Games are severely underrepresented in both respects; while games have several characteristics that cater directly to neurodivergent sociality, they are rarely studied in the social computing context. For further analysis and descriptions of each research area, see Section 5.

\subsection{Demographic Insights (RQ1)}

\begin{table}
\small
        \centering
        \begin{tabular}{l|r|r|r}
        \toprule
         \textbf{Neurodiverse Focus} & \textbf{2013--2018} & \textbf{2019--2023} & \textbf{Percent Change} \\ \midrule
         ADHD & 2.56\% & 11.11\% & +8.55\% \\ \hline
         Dyslexia & 2.56\% & 8.89\% & +6.32\% \\ \hline
         Learning Disability & 2.56\% & 8.89\% & +6.32\% \\ \hline
         Intellectual Disability & 12.82\% & 17.78\% & +4.96\% \\ \hline
         Cognitive Disability or Impairment\footnote{While the use of the word "impairment" is recognized to be harmful, some papers in our corpus used this term as a classification for neurodiverse participants.} & 7.69\% & 6.67\% & -1.03\% \\ \hline
         Asperger's Syndrome & 5.13\% & 0.00\% & -5.13\% \\ \hline
         Autism & 61.54\% & 55.56\% & -5.98\% \\ \hline
         Cerebral Palsy & 17.95\% & 4.44\% & -13.50\% \\ \bottomrule
        \end{tabular}
        \caption{Share of papers in the corpus that mention participants with the listed neurodivergent identity. Note that the appearance of an identifier in this table does not indicate the authors of this paper believe the term is appropriate. See Section 8.2.5 in the Appendix for descriptions of each identity.}
        \label{ndfocus}
    \end{table}

\textbf{A majority of ND social computing research focuses on autism, likely because it is well known and involves speech practices that sometimes differ from those of neurotypicals.} A sizable portion of the papers selected (n=49, 58.33\%) include autistic participants, with 38 focusing exclusively on autistic participants. This is expected, as most studies of neurodiversity often feature primarily or exclusively autistic participants, and autism is one of the more widely known neurodivergent identities. Additionally, funding is more readily available for studies with autistic participants as compared to other ND identities \cite{williams_perseverations_2020}. 
ADHD and Dyslexia research are becoming increasingly common. However, ADHD is rarely the primary focus of a project in our corpus. While co-occurrence with autism is common and should be studied, the ADHD-specific needs of users are often overshadowed by characterizations that can be applied to the broader population of autistic participants.
For example, ADHDers and their communities appreciate shared reflection but struggle to integrate this practice into daily life \cite{stefanidi_children_2023}. Attention to the multimodal literacies of dyslexic people on SNS offers new ways for dyslexic people to engage socially using mediums other than text \cite{vezzoli_dyslexia_2017}. The integration of these features necessitates a diversity of neurodivergent people, including autistic people, ADHDers, dyslexic people, and still others.

\begin{figure}
    \centering
    \includegraphics[scale=0.7]{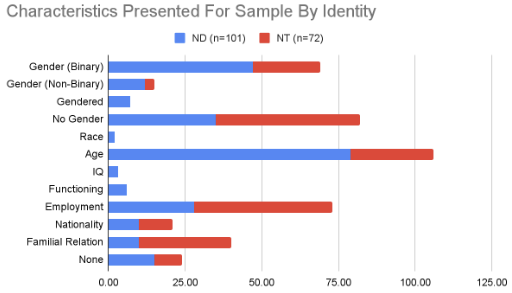}
    \caption{Identifying characteristics presented for each of the 101 neurodivergent and 72 neurotypical samples. A sample could be identified across multiple characteristics (ex. a description of "children ages 7-17 from the United States" would be coded under "Age" and "Nationality"). ND participants are most often described by their age and binary gender (male or female) when gender is presented. NT participants are most often identified exclusively by their familial relation to a corresponding ND participant or their employment (teacher, therapist, etc). For more information on the application of identifying characteristics, see "Identifiers" in Section 8.2.8 of the Appendix.}
    \Description[Neurodivergent participants are most often identified by age and binary gender, whereas neurotypical participants are identified by their employment and gender is not stated.]{Of the 101 samples of neurodivergent people, the following number of samples present the following characteristics: 47 by binary gender, 12 by non-binary gender, 7 by a gendered pseudonym, 35 with no stated gender, 2 by race, 79 by age, 3 by IQ, 6 by functioning level (high or low), 28 by employment status, 10 by nationality, 10 by familial relation, and 15 had no identifying characteristics. Of the 72 neurotypical samples: 22 by binary gender, 3 by non-binary gender, 0 by a gendered pseudonym, 47 with no stated gender, 0 by race, 27 by age, 0 by IQ, 0 by functioning level (high or low), 45 by employment status, 11 by nationality, 30 by familial relation, and 9 had no identifying characteristics.}
    \label{charspresented}
\end{figure}

\textbf{ND contributions are almost always accompanied by (or excluded in favor of) NT participation.} The average number of neurodivergent participants per sample was 15.60 and the average number of neurotypical participants per sample was 29.71, while the median for each group was eight (below the median of 13 found by Mack et. al.'s investigation of accessibility research \cite{mack_what_2021}). This suggests that, for every experimental setup that consults ND and NT participants, there are approximately two NT participants for each ND participant. Furthermore, 63.1\% of papers (n=53) include participants that are presented as (or implied to be) neurotypical, whether it be medical experts, parents, or neurotypical children. This is much higher than the standard inclusion for non-disabled or expert participants in accessibility research \cite{mack_what_2021}, suggesting that ND social computing researchers heavily value neurotypical comments, even in a field that warns against speaking for instead of with disabled people \cite{rogers_does_2013}.
\begin{figure}
    \centering
    \includegraphics[scale=0.7]{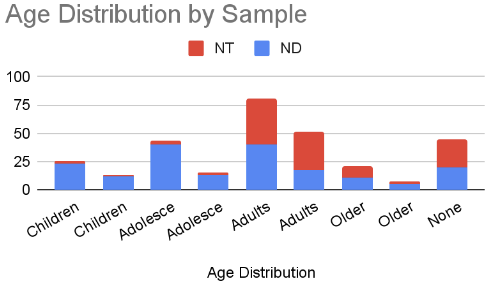}
    \caption{Age distribution by sample type. Note that the number of samples does not sum to 181 because multiple age groups can be present within a sample and some samples do not distinguish between ND and NT participants. NT involvement in ND social computing research is almost exclusively restricted to NT adults, except for a few cases in which ND and NT children are directly compared. }
    \Description[Neurodivergent participants are almost always under the age of twenty, whereas neurotypical participants are almost exclusively adults over the age of twenty.]{Of the 101 samples of neurodivergent people, the following number of samples include the following age groups: 23 include children aged 12 and under (12 only include children), 40 include adolescents aged 13-19 (13 include only adolescents), 40 include adults aged 20 to 49 (18 include only adults), 11 include adults older than 50 (5 include only older adults), and 20 have no ages listed. Of the 72 neurotypical samples: 2 include children aged 12 and under (1 only includes children), 3 include adolescents aged 13-19 (2 include only adolescents), 41 include adults aged 20 to 49 (33 include only adults), 10 include adults older than 50 (2 include only older adults), and 25 have no ages listed.}
    \label{agedist}
\end{figure}

\textbf{NT participants are almost always adults paired with children.} A potential explanation for the abundance of NT participants in ND social computing research is the involvement of parents in research with ND children. As such, in studies with both NT and ND participants, NT participants are almost always older than ND participants. With this age difference comes additional authority, as these older participants are often caretakers or parents of the ND children participating in the study. This heavily influences the platform dynamics of the resultant systems.

\textbf{ND social computing study participants are slightly older than ND participants from neurodiversity research at large.} The age of our participants is slightly older than previous reviews of neurodiverse HCI research \cite{spiel_purpose_2021, spiel_adhd_2022}. Most likely this is because the kinds of open-ended systems that our review focuses on are often deemed inappropriate for children under the age of 13. What is more surprising is the massive range of several samples; several papers operate on the Australian definition of “young adult” (18--34 years old), and others include children and adults who have vastly different communication capacities and needs \cite{narain_personalized_2020}. ND adolescents (13--19 years old) are rarely studied on their own; usually samples will also include children younger than ten or adults in their late twenties in samples that include teens. Similar to NT teens, ND teens are highly engaged with social media, struggle with presentation online, and heavily rely on social platforms to connect with peers \cite{rocheleau_privacy_2022, vezzoli_dyslexia_2017}.

\begin{figure}
    \centering
    \includegraphics[scale=1.1]{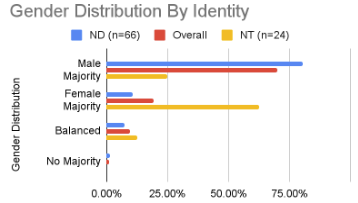}
    \caption{Gender distribution by identity. A male majority sample is a sample of n participants where at least (n/2) + 1 participants are male, a female majority sample has (n/2) + 1 female participants, a balanced sample has the same number of male and female participants, and a sample with no majority features no group with more than n/2 participants. ND samples are much more likely to be male majority, and NT samples are much more likely to be female majority.}
    \Description[Of the samples that disclose the gender of each participant, neurodivergent samples are far more likely to have more male participants whereas neurotypical samples are more likely to have female participants.]{Of the 90 total samples that specify the gender of each participant, the classification of each sample based on gender is: 65.56\% male majority, 24.44\% female majority, 8.89\% balanced, and 1.11\% no majority. Of the 66 neurodivergent samples: 80.30\% male majority, 10.61\% female majority, 7.58\% balanced, 1.52\% no majority. Of the 24 neurotypical samples: 25.00\% male majority, 62.50\% female majority, 12.50\% balanced, and 0.00\% no majority.}
    \label{genderdist}
\end{figure}

\textbf{ND samples are almost exclusively male, and NT samples are mostly female.} Of the 66 samples with ND participants that provide the gender of each participant in the sample, a large number of these samples are majority male (80.30\%), meaning there are more male identifying participants in the sample than any other gender identity. Males are more likely to be diagnosed with a neurodivergent condition than females (despite similar numbers of ND adults \cite{fairman_diagnosis_2020}), so this disparity seems consistent with a medical framing of neurodivergence. What was most surprising, however, is that neurotypical samples are often majority female (62.50\%). This suggests that projects focusing on relationships between neurodivergent and neurotypical people are most often considering only one type of relationship: not just parent and child, but mother and son. This is further evidence of the gendered expectations for caretaking and emotional labor in ND research, which could be worsened by normative technologies that expect mothers to learn to use new and complex systems \cite{ms_gendered_1994}.

\textbf{ND social computing research primarily presents binary gender.} Neurodiverse and LGBT communities often collaborate and have concurrent members \cite{miller_i_2020}, meaning it is important for researchers to allow neurodivergent participants to present themselves as they identify to construct a representative sample. Our corpus suggests that research in this domain is not inclusive; most samples identify participants in a binary manner (ND 46.53\%, NT 30.56\%, Overall 38.12\%) or do not present gender (ND 34.65\%, NT 65.28\%, Overall 46.96\%). Few papers allowed participants to disclose as non-binary or choose not to disclose their gender (ND 11.88\%, NT 4.17\%, Overall 11.05\%). Although this may be an improvement over Spiel et al.’s finding that autistic children are strictly presented in binary terms \cite{spiel_agency_2019}, diversity in representation is far from widespread. More inclusive language in DSM-V-TR \cite{association_diagnostic_2022} and more gender options in recent papers suggest that gender identification practices are changing to be more inclusive. Open-ended identification options and recognition of LGBT ND people is becoming increasingly common \cite{choi_love_2023}.

\textbf{Formal diagnosis is usually a requirement for participant inclusion, and self-identification alone does not lead to more representative samples.} A slight majority of ND samples (n=52, or 51.49\%) explicitly require a formal diagnosis to take part in the study, and others do not mention whether diagnosis was used in sourcing participants (n=17, or 16.83\%). As stated earlier, males tend to be formally diagnosed as neurodivergent at a younger age despite similar numbers of ND people by gender in the adult population \cite{fairman_diagnosis_2020}. 
However, allowing participants to self-identify and forgo formal diagnosis may not be enough to facilitate equitable research. Of the 17 samples that allowed self-diagnosis, 11 were majority male (64.71\%), four were majority female (23.53\%), one was balanced, and one did not specify the gender distribution. While this is less extreme than the 80\% male majority described above, many of these samples are still heavily male (including male to female ratios of 9:1 \cite{morris_understanding_2015} and 5:1 \cite{choi_love_2023}).

\subsection{Language and Framing (RQ2)}
\textbf{Positionality statements are rare and are almost exclusively included by ND researchers \cite{liang_embracing_2021}.} Positionality statements are practically nonexistent in our corpus; only 6 papers (7.14\%) address their relationship with the neurodiverse population of study, and practically all of these are to disclose that their own identity matches those whom they study (n=5, 5.95\%). 
Some types of disclosures present in our corpus included: disclosing at least one member of the research team is neurodivergent, acknowledging consultation with neurodivergent people, and stating the educational background of the researchers and its influence on the project.

\textbf{While many authors do not explicitly adopt a medical model of disability, much of ND social computing research is highly medicalized.} While coding for the perspective authors take toward neurodiversity in our corpus, we struggled to classify each paper as adopting a medical model, social model, or otherwise. Thus, we instead coded for motivations for constructing or designing the system of interest. The full list of attributed labels is located in the Appendix in Section 8.2.6. Many papers had a stated goal of improving productivity (n=17, 20.24\%) or mitigating negative symptoms (n=16, 19.05\%). The language used in these papers is consistent with a medical model of disability, using almost exclusively negative language to describe characteristics of neurodivergent people and framing technology as a mitigating agent. Similarly, many papers that identified strengthening relationships involving ND people as their goal (n=13, 15.48\%) framed expressions of neurodiversity as negatively impacting relationships with family and friends. 
While a medical approach in social systems can be appropriate, as in the case of preventing self harm and promoting mental health \cite{jo_geniauti_2022},  it is often inappropriately invoked to legitimize normative conditioning, examples of which are described in the following section. The perspectives of understanding existing social practices (n=23, 27.38\%) and the non-prescriptive augmentation of ND speech (n=15, 17.86\%), which are most aligned with the social model of disability, are most often used by authors that have multiple papers in our corpus or include positionality statements that disclose their neurodiversity. There is no clear trend in the usage of these different perspectives across time.

\textbf{Verbal ability and intelligence are often conflated in ND social computing works that adopt medicalized perspectives.} Disabled scholars have described how harmful it can be to accentuate the ``problem'' of disability, as is common in medicalized approaches to research with disabled participants \cite{ymous_i_2020}. Descriptions of deficits are discouraged by language guides like those provided by ASSETS \cite{hanson_writing_2015} but are distributed throughout the literature. While neurodivergent people do experience difficulties acting within neurotypical institutions, these descriptions place blame on the individual for being confused or incapable. Some papers outright question ND users’ ability to contribute meaningful feedback with references to IQ and verbal ability. Conflation of verbal ability and intelligence among researchers and the general public can lead to material harm against neurodivergent people, as was the case in the following paper from our corpus:

\begin{quote}
    “G4-T reported that she used to discipline the child for making noises. However, using TalkingBoogie-Coach, she began to observe the intent of the child in order to share it, and found out that the child tried to talk about TV programs by drawing attention.” \cite{shin_talkingboogie_2020}
\end{quote}

Normative assessments like these (“noises are undesirable”) can have damaging effects (“neurodivergent speech is intolerable”). This intelligence-based framing can also lead to prescriptive rather than descriptive insights, ignoring material problems in favor of actionable “solutions” generated by “expert” requirements. 

\textbf{ND participants are often portrayed as untrustworthy, and even open-ended systems have implicit goals.} The rhetoric of burdens and antagonism are also relatively common. A project that develops a social media application for people with intellectual disabilities defines “learned helplessness” as a behavior that the platform seeks to eliminate \cite{xu_rove_2014}, which evokes the psychiatric paradigm of “escape to health” that positions researchers in opposition to opportunistic behavior \cite{edelman_political_1974}. Unexpected behaviors are almost always viewed negatively, even though they can be sources of resistance and innovation that allow neurodivergent participants to express social values that cannot be supported by normative systems.




\subsection{Participation and Choice (RQ3)}

\textbf{ND participants are almost always involved in the evaluation of systems, but rarely consulted in earlier phases of the research process.} Each paper in the corpus was coded for the involvement of ND participants in the requirements, design, and evaluation phases of the development of technology. In our corpus, only 13\% (n=11) of papers were fully participatory, meaning ND users were asked what functionality the technology should have, given opportunities to revise the structure of the system during its development, and evaluated whether the system performed its designated functions \cite{corlu_involving_2017}. This is consistent with Mack et al.’s assessment of the prevalence of fully participatory design in accessibility research generally \cite{mack_what_2021}. Most surprising is that 56\% (n=47) of papers only involve ND users in the evaluation phase, and 20\% (n=17) do not involve ND people directly in their project at all despite them being the primary stakeholder in the system. The implication is clear; in ND social computing research, ND people are most often subjects only included to test if a device “works” rather than participants in the research process.

\textbf{Despite the stated goal of developing open-ended systems, the modes of communication that researchers use further limit the ability of participants to define social success in their own terms.} Of the communication mediums used in the 181 samples within our corpus, observational studies were the most common (n=56, 30.94\%), followed by in-person interviews (n=37, 20.44\%) and workshops (n=27, 14.92\%). To emphasize the tendency toward involving ND people only in evaluation, 23.20\% (n=42) represented observational studies in which the observation (often presented as user data independent of the experiences of the ND evaluator) was the only way in which ND people were involved in the project. Despite their rarity relative to interviews (in person and virtual) and observational studies, workshops were successful and highly praised by participants when attempted, despite some initial preconceptions revealed by a particular project:

\begin{quote}
    “Participants shared that they felt safe and comfortable... From our observations, participants showed no sign of awkwardness despite their ASD condition even in the first workshop. Participants were able to engage in small talk with each other during the workshops. One participant even brought homemade snacks to share with the design team.” \cite{zhu_co-designing_2019}
\end{quote}

This is another reminder that the only way to know the social needs of a neurodivergent participant is to ask. Many researchers make assumptions about what conditions would best facilitate neurodivergent participation (e.g., including parents in interviews, conducting interviews over text chat) when neurodivergent populations can be diverse and include different preferences even within the same diagnostic criteria. 
Unfortunately, few projects in the corpus allowed participants to choose their communication medium in interviews \cite{race_understanding_2022, dai_designing_2022, page_perceiving_2022}. Some conducted interviews across mediums, but only offered remote options due to physical constraints rather than communication preference \cite{morris_understanding_2015}. Variable involvement of parents or caregivers in the interview process is incredibly rare; only one paper explicitly mentions ND participants were given the opportunity to conduct the interview with or without a parent \cite{rocheleau_privacy_2022}. 

\textbf{Compensation, a vital factor that defines a participant’s willingness and ability to participate in research, is rarely mentioned.} Besides the medium through which a participant can engage with a research process, arguably the most important factor determining participation is compensation. Despite calls to standardize the reporting of compensation in HCI research \cite{pater_standardizing_2021}, compensation amounts are so rarely reported that the authors were not able to code for them in our corpus.
Especially when scraping sensitive personal data on public forums without consent, as some papers in our corpus do, researchers may even cause harm by presenting identifiable characteristics that they believe to be properly anonymized. Research undertaken without contacting participants risks the proliferation of conclusions that are largely unfounded \cite{shapiro_viral_2020}, besides the harm caused by the use of this information being uncompensated.


\section{Results By Research Area}

While all of these technologies fit our inclusion criteria, they represent radically different systems with unique features and use cases. We now analyze each research area for methods, trends, and design recommendations (RQ4). By developing a deeper understanding of methods by which each system is studied, we can develop domain-specific recommendations for designing neurodiverse, accessible systems.

\subsection{Social Network Services (SNS)}
The most popular systems studied were social network services, or SNS, with 27 papers studying existing platforms or creating new ones. 
Although research before 2020 predominantly focuses on the creation of new platforms exclusively for neurodivergent people and caretaker experiences, the last three years have focused on existing community spaces and cooperative sense-making tools. The progression of this area suggests that researchers may be adopting a stronger interdependence-conscious framing for neurodiverse social media that understands the importance of community for lateral rather than top-down support structures \cite{bennett_interdependence_2018}.

\textbf{Neurodivergent people are often considered consumers rather than creators of content on SNS systems}. Mok et al.’s study of autistic livestreamers \cite{mok_social_2022} and a set of papers on dyslexic content creation \cite{vezzoli_dyslexia_2017, wu_design_2019} are the only SNS papers that discuss neurodiverse creators as generating content to socialize with an audience that is not exclusively ND people and their caretakers. There are more articles that mention the practices of caretakers (autism bloggers, tech caregiving) than community building. By neglecting ND people as content creators, researchers  conflate difficulties with being social and an aversion to being social. Pena et al.’s work with autistic adults is a prime example of ND people who have complex curation processes so they can engage with others: difficulties with phatic conversation, highly specific filtering needs, and time management strategies beyond time limits increase the work required to create the content they would like to see on SNS \cite{barros_pena_my_2023}. Rather than removing choice to encourage simplistic interactions, future ND social computing research should explore how to make feed curation and content creation a transparent, collaborative process between ND creators and their communities. ND creators in SNS
have complex invisible labor practices, and this effort could be lessened through the formalization of these practices in ND social computing systems \cite{star_layers_1999}. Tools that facilitate this work would promote enjoyable, reaffirming, and fulfilling experiences for a key constituency of SNS platforms.

\textbf{Privacy and safety is the most prevalent research focus among SNS ND social computing researchers.} Much of this is informed by parent and caretaker concerns, as well as medicalized research guidelines designed to “protect” neurodivergent participants. However, the motivation to restrict ND expression for the sake of safety is strongly tied to recruitment methods. Studies of social media behavior that are recruited from skills training centers provide detailed narratives of participants doxxing themselves and being reliant on peers for validation \cite{page_perceiving_2022, bayor_characterizing_2018}. These studies recommend defensive threat models that ensure neurodivergent users do not inadvertently expose themselves to violence, online or otherwise. Consider the following study that recruits from several such locations and the resulting generalization that is applied to all people with cognitive disabilities:

\begin{quote}
    “The selected institutions are 1) a primary school for youth with autism, 2) an education programme for youth with special needs and 3) a sheltered home for youth with cognitive disabilities. \textit{The institutions represent three stages that young people with cognitive disabilities typically go through in their life} (emphasis added).” \cite{andreasen_digital_2019}
\end{quote}

While possibly appropriate for ND people who engage with institutions similar to those above, a system designed for participants recruited in this way is not necessarily useful to a representative sample of ND people, many of whom do not or are unable to access these services. Studies that recruit from online or decentralized sources find ND people are more cautious with their SNS use than their NT counterparts. For example, one study assumes that a set of autistic users that were recruited online are more vulnerable to phishing attacks, but finds them more capable than neurotypical users at finding malicious links \cite{neupane_social_2018}. A project with a similar recruiting scheme studying privacy perceptions of autistic young adults finds them more risk-averse on social media than neurotypicals of the same age group \cite{rocheleau_privacy_2022}. ND social computing work on SNS needs to recognize these conflicting beliefs about privacy are not monolithic and are heavily dependent on context and individual preferences. There are ND people that do not share the concerns of their peers about online threats, and there are ND people that have trauma from online harassment. In either case, risk taking is a part of self-disclosure and vulnerability which is how people find community. Future ND social computing work should recognize this tension and how ND people can work with different types of communities to feel supported online. Recruitment methods that are more conscious of the differences between these groups (e.g., a skill training center for autistic teens, autistic self-advocates on Facebook, or some combination of these groups) would be instrumental in developing the specifications needed to construct safety features that address the needs of ND people across large diagnostic categories.

SNS are open-ended social environments that present plenty of risks, to the degree that parents and researchers deem simplistic and isolated platforms preferable to the open internet. However, as Rochleau and Chiasson and many others have shown \cite{rocheleau_privacy_2022}, ND people can also be conscious of these risk factors and form online communities with the infrastructure to protect themselves and others. Rather than reducing options for ND people as a protective measure, there is an opportunity for first and third party systems to offer transparent and collaborative content curation solutions that assist ND users in the complex curation practices they have already developed.

\subsection{Augmentative and Alternative Communication}
Augmentative and Alternative Communication (AAC) systems are most commonly used by nonverbal autistic and physically disabled people to verbally communicate by directly translating non-verbal inputs into verbal speech. 
AACs present an opportunity for nonverbal people to communicate in settings that would otherwise be restricted to verbal communications, such as in person service interactions, potentially giving ND people increased agency in their daily lives.

As an example of an AAC system, the ECHOS project from MIT features several questionable design choices that interrogate the values of AAC research. Rather than a traditional AAC system, researchers attempt to build a suite of tools that classify and track the behaviors of nonverbal autistic youth to allow for direct nonverbal communication between children and their parents. Although the following quote is not from the ECHOS paper that is included in our corpus\footnote{Note that we excluded multiple publications with the same sample and concerning the same system in our search. Representation of the ECHOS project in our corpus consists of a later publication \cite{narain_personalized_2020} with more details concerning participants and their involvement with the project.}, it is the most prominently featured output from the project and is demonstrative of its values:

\begin{quote}
    “While the sensors themselves did not appear to be irritating, the gel electrodes, particularly from the ECG Biopatch sensor, left a sticky residue on the skin that was difficult to remove… which led to mild skin irritation. In response, a cloth-based Bioharness (Medtronic, USA) was trialed, \textit{but the data quality and fit was poor}…. In addition, the requirement of having the researcher present to attach the sensors – and the sensors being too expensive to leave with the participant – meant that \textit{this setup would not be able to reach the geographically distributed participants who could benefit from this technology} (emphasis added).” \cite{johnson_echos_2020}
\end{quote}

There are two underlying assumptions here that take AAC research further away from the interests of ND users. The first is that researchers need high quality usage data to justify their project, which often involves invasive tracking of ND speech. The second is that the parents are the anticipated customers that these products are targeted toward, as they are the ones who often purchase these systems for their children. This can be clearly parsed in the later portion of the above quote: painful skin irritation is not a disqualifying factor, yet the inability to market and distribute the system dictates what sensor should be used. Additionally, if parents control the construction of the vocabulary of the system (which is often the case), this could limit the ND user’s expression to the completion of tasks that the parent deems acceptable. \textbf{Rather than providing ND people with more options, AACs can be misused such that they further limit ND expression and cause pain (either mentally through normative conditioning or physically due to uncomfortable, invasive equipment).} 

Future research could learn from projects like TalkingBoogie \cite{shin_talkingboogie_2020}, which designed an AAC in a way that integrated both parents and children in critically analyzing the shortcomings of the system. By developing both an AAC application and a separate application that allows parents and teachers to notify each other and monitor issues with the system, the AAC's interface can be customized to better meet the user's needs. This collaborative structure is supported by Valencia et al.'s research, which shows that conversation partners of people using AAC systems often try to be helpful by predicting the speech of or talking over an AAC user, but this often leads to frustration as AAC users are often unable to correct or comment on topics before the conversation has moved on. Future systems should consider features like those of TalkingBoogie that promote conversational agency by ensuring verbal conversational partners respect the time and privacy of the user as they construct an utterance \cite{valencia_conversational_2020}. This prompts the following question in relation to the application of AI and sensors that predict ND speech: instead of speeding up ND speech with predictive algorithms and sensors that speak for ND people like an interrupting conversation partner, how could an AAC be designed such that a conversation partner allows the user to complete the construction of their utterance? This echoes our call for double empathy, as it promotes a consciousness of both sides of the conversation rather than engineering systems only on the ND side \cite{milton_ontological_2012}. \textbf{Both members of the conversation, the person directly interfacing with the AAC and the conversation partner, must be accounted for to ensure AAC users have the agency they deserve as they negotiate their social experience.}

\subsection{Digital Collaboration}
Digital collaboration systems provide a technical context through which information can be shared between users collaborating on a shared desired outcome. Some examples of such systems in our corpus include: empirical studies of teleconferencing \cite{mokashi_exploration_2013, zolyomi_managing_2019, tang_understanding_2021, das_towards_2021}, assistive conversational interfaces \cite{yaghoubzadeh_spoken_2015}, and wearable communication aides facilitating asynchronous, non-semantic communication \cite{simm_prototyping_2014, silva_unpacking_2023}. Because these systems are often used in work and school environments, researchers often use centralized recruiting schemes (schools for children with learning disabilities, skills training centers for young adults, etc.) to find participants. Similar to what was discussed in Section 5.1, implications from these papers are heavily influenced by a recruitment scheme that privileges caregivers or staff who value security over self expression. 

ND involvement in digital collaboration systems research varies greatly: some involve exclusively ND people \cite{zolyomi_managing_2019}, others attempt to balance the needs of ND and NT collaborators by discussing each person’s needs individually \cite{mokashi_exploration_2013}, and still others exclude ND participants entirely despite stating that ND people are imagined users of the system \cite{corlu_mediated_2016}. Furthermore, much of workplace research is dominated by a stereotypical NT manager to ND employee, top-down collaboration structure. As most collaborative spaces contain differing expressions of neurodiversity, it is important for these systems to address the needs of ND people as they coordinate with others (allowing for multimodal communication, providing materials in advance of meetings, etc.) and the process by which these needs are met by both ND and NT collaborators in a more lateral manner \cite{das_towards_2021, simm_prototyping_2014}.

\textbf{Personal and family informatics present a promising opportunity to allow neurodivergent users to exercise executive function by logging and selectively sharing information to collaborate with friends and family \cite{pina_personal_2017}.} The Clasp \cite{simm_prototyping_2014} project turns a stress-ball into a customizable communication tool with programmable message-sending capabilities in periods of high stress. Rather than translating input to literal speech like an AAC, Clasp allows ND users who may struggle to construct utterances in periods of high stress to squeeze the ball in order to send a pre-written message to a trusted person. CoolTaco allows ADHD children and their parents to collaboratively set goals and document routines \cite{silva_unpacking_2023}. Both systems are not without flaws, as they are implemented in a way that promotes reliance on an implied neurotypical authority (e.g., the NT person “saves” the ND person in distress, the NT parent “rewards” the ND child for ``good'' behavior). 
However, their apparent success suggests that ND users could benefit from wearable technology, as long as they can achieve control over their personal data. Taking inspiration from trauma-informed design \cite{scott_trauma-informed_2023}, future research should explore how ND people receive peer support online in times of crisis, select information to share, and voice concerns across various communication mediums (e.g., tactile, vocal, text) to find new ways to work with peers in triggering situations.

\textbf{Rather than exploring unique forms of input, most papers focus on the pros and cons of teleconferencing for neurodivergent students, professionals, and older adults.} This renders many digital collaboration systems useless for ND people who exist outside of places like skill training centers or tech companies. The prioritization of “productivity” almost always excludes people who are nonverbal, unable to work in a traditional office space, or cannot work regular hours. This bias is supported by mentions of “high functioning” and “low functioning” people in discussions of autism, which are more frequent in papers of this research area. These terms are problematic on their own; being labeled “high-functioning” often belittles the difficulties of autistic people in the workplace, while being labeled “low-functioning” is an assessment often based on verbal speech, ignoring other literacies or intelligences \cite{williams_fallacy_2019}. Rather than prioritizing vague definitions of functioning over the lived experiences of nonverbal and minimally verbal people in the workplace, we challenge researchers to conceptualize ways to communicate in meetings beyond the audiovisual mediums that are presently available. Future ND social computing work should adopt a holistic approach that investigates technology-assisted work of ND people in non-technical environments and across a variety of verbal preferences. Digital collaboration systems are often used to stay connected with friends and family both for daily work and in times of crisis, meaning it is essential that these systems are inclusive of a variety of modes of communication.
 
\subsection{Sensors for Social Feedback}
Social feedback systems in our corpus are typically visual aides that translate latent social information from the ND user's environment into digital cues. If the AAC systems are primarily designed to help ND users externalize or output communication, social feedback systems intake and display ambient information to ND users to aid in conversation. Common types of social feedback systems are facial expression monitoring \cite{lin_empathics_2020} and distance measurement for conversation partners \cite{halperin_social_2016, boyd_procom_2017, crowell_variations_2018}. We note that our corpus excludes the many social skills training platforms in our search because they do not feature open-ended conversation partners---most facilitated existing structured therapy regimens. Those included in our corpus are conversational aides that are embedded into heads-up-displays or other devices and are intended to be used in public settings.

Because these technologies are meant for use in public life and are highly visible, it is worth noting that interaction between the visibility of assistive technologies and the visibility of disability is a complex, personal process \cite{faucett_visibility_2017}. For example, a blind person may use a white cane for navigation or they may use a navigation app, depending on the affordances of each system and perceived stigma. Because neurodivergence is often visible within social interactions as it often directly impacts speech, a variety of high visibility and low visibility technology could give neurodivergent users choices in self presentation that would otherwise be unavailable. With AACs being highly visible communication aids, it would be reasonable to expect these social feedback systems to offer more subtle assistance. However,  many of the systems in this category have high visibility, offer little customization to augment this visibility, and encode normative values without considering the social aspirations of the user. The lack of customization is further explained by the lack of specificity with which prospective users are defined. One device claims to be used by “people who are blind or have social disabilities” \cite{halperin_social_2016} while another is for “doctors and autistic adults”, claiming both have issues with empathizing \cite{lin_empathics_2020}. Those that are designed solely for neurodivergent users are usually targeted for children to learn social norms in public settings. While researchers often claim that their systems are non-prescriptive and are therefore immune to the criticisms of normative therapy schemes, this assertion is often negligent of the reality of their use cases. Consider SayWAT, a device meant to alert autistic adults of atypical prosody (tone) and inappropriately high volume. The system is portrayed as non-normative, as it objectively reads ambient noise and displays notifications once a defined threshold is reached. A hypothetical use case, however, indicates a clear normative goal. The description of this use case involves an autistic college student, Jane, meeting her roommate, Lucy, for the first time and is concluded in the following manner:

\begin{quote}
    “Jane decides to check in on her pitch, knowing she can tend to sound a bit bored, and she wants Lucy to know how excited she is… After a few alerts, she successfully modulates her voice, and Lucy, noticing the change in her emotion, smiles at her new friend.” \cite{boyd_saywat_2016}
\end{quote}

The implication is clear: Lucy would not like Jane without the device. There is no discussion of how rude Lucy is, or whether she is maintaining proper eye contact, distance, or tone during this conversation. It is solely Jane that must perform, and she must follow the device's instructions to do so. Participants themselves point out this asymmetry, but a significant portion of the limitations section is instead dedicated to the problem that “an individual with autism may not be aware of atypical prosody much less be seeking a solution for it” and how this dissonance could be resolved with savvy marketing.

\textbf{ND end users are often not the primary stakeholders of these technologies, as their objective is often to reduce the discomfort of the NT stakeholders surrounding them.} These technologies are typically made to placate neurotypical misunderstandings of and reduce discomfort around ND expression. Consider two more examples, the first of which is a robot companion that assists and converses with an autistic child throughout their day, then delivers an assessment of what “percent autistic” a child is on that given day \cite{chen_cognitive_2021}. What is presented as a conversational aid and companion for a child is in reality a surveillance tool that encourages parents to reward behaviors that lower an arbitrary “autism score”. This is a rehash of anti-autism sentiment rebuked by Sinclair over thirty years ago, but with the added pseudo-scientific credence of an algorithm. As such, the child-facing features are minimally discussed in favor of the atypical-expression-detection model used to assign the child’s score. Whereas Section 5.3 exposed the potential of wearables to increase ND agency, we found that social feedback systems often result in invasions of privacy and can even potentially endanger ND users. For another example, consider ProCom, a distance measurement device for autistic users to keep proper space between them and their conversation partner \cite{boyd_procom_2017}. Only parents and medical experts are consulted in the design phase, which focuses on the inability of autistic children to gauge distance between conversation partners. The only evaluation for the device given by the children are the following:

\begin{quote}
    “it was really cool but creepy how it knows where you are.” (P3, 12-year-old boy)
\end{quote}
\begin{quote}
    “If you made like a decorative belt and made little dots this size, it would actually look like a belt so it doesn’t look like you’re spying on everybody” (P7, 11 year-old girl) \cite{boyd_procom_2017}
\end{quote}

Somewhat ironically, the primary concern of the parents was that their children put themselves in compromising social situations. The evaluation of the success of the technology in this section is focused exclusively on parents’ comfort and wishes for their children, yet the children themselves are more conscious of how these devices may be viewed negatively by others. The first author of both SayWAT and ProCom has discussed problematic features of these systems in detail \cite{williams_counterventions_2023} and offers suggestions toward creating useful social feedback systems with and for ND users. Future ND social computing research should be cautious in using sensors and ambient displays for several reasons. \textbf{Principally among them is the inability to make a truly non-prescriptive social feedback system, as ``translating'' social information necessitates some form of standardization or interpretation.} Although several papers claim to not make normative judgements, it is impossible not to internalize the directions of red text saying “too close” or a blue frowny face labeled “sad”. Instead of designing systems that place the responsibility of maintaining conversational norms entirely on the ND person, researchers should examine existing needs of users and the conversation partners that neglect these needs. As seen in the examples above, devices produced without an awareness of double empathy risk making social interaction even more demoralizing for ND users.

\subsection{Workplace Culture and Organizations}
This set of systems in our corpus includes studies that, while they may concern systems that could be placed under one of the prior categories, often include an ecology of many such sub-systems embedded in workplaces or experiential programs where ND people are the primary users. These papers focus not only on the technologies used by ND people, but also the social dynamics and contexts in which these technologies are learned and used. Technologies discussed include a combination of teleconferencing, collaborative IDEs, 3D printing, AR/VR experiences, and others. All of this work is empirical, and nearly half of the papers focus on the experiences of support staff and caretakers rather than ND people \cite{sharma_socio-technical_2020, balasuriya_support_2022, wang_virtual_2023}. Similar to the Digital Collaboration section, depictions of ND people are restricted to those in higher education or professional settings \cite{morris_understanding_2015, zolyomi_values_2018}, whereas papers focusing on support facilities only conference with staff in skill training centers and parents. There is also a set of papers that focus on summer camps for ND children \cite{moster_can_2022, wang_virtual_2023} that rely  on staff narratives of the experience rather than the perspectives of the ND participants. Thus, what we know about ND people in these facilities and programs often is only available as a third-person narrative rather than a first-person account.

That being said, these papers highlight that: \textbf{many NT people are implied users of ND social computing systems who have needs of their own that must be captured in the design of ND social computing systems.} Support staff are also users of many of the technologies described in our review, and Balasuriya et al.’s work provides valuable insight into the centers that several papers in our corpus recruit from. Many staff emphasize insurance and regulatory requirements within these programs; if participants are not learning tangible skills that are valuable for employment, the center risks losing funding. Staff are overwhelmed by the constant introduction of new technology, and as a result experiences offered to members are heavily dependent on the interests and abilities of the staff \cite{balasuriya_support_2022}. This context further emphasizes how essential it is to understand the environment within which these technologies are used, as complex technologies that do not meet regulatory requirements will not be maintained even if they facilitate positive experiences. The influence of insurance in these facilities also emphasizes the biases inherent in recruiting from these facilities, as this sample is comprised of those who are insured and financially supported. This also explains the many papers found in our search that followed defined therapy regimes, as technologies that follow these strict guidelines may be seen as potentially profitable investments despite their potential for harm through normative conditioning \cite{kupferstein_evidence_2018}.

\textbf{Irrespective of the involvement of ND people, the relative successes and failures of workplace environments and skills training programs are often conceptualized with respect to ND productivity and employment.} Studies of ND people in the workplace are often of ND computer scientists in industry and academia. However, there are many ND people that work with social computing systems outside of the tech industry and have lower tech literacy that are excluded from these conversations of inclusive workplaces. And again, we found none of these systems consider how NT people can better support ND people through double empathy \cite{kim_workplace_2022}. 
Future research should consider the workplace in the context of double empathy and design for ND productivity with an awareness of NT engagement with ND working styles.

\subsection{Games}
Games in our corpus take the form of digital worlds that allow users to collaborate, compete, and communicate in highly structured, game-specific architectures. Five out of the seven games in our corpus are ethnographic studies of the Minecraft server Autcraft by Ringland \cite{ringland_making_2015, ringland_would_2016, ringland_will_2016, ringland_making_2017, ringland_place_2019}. The remaining two are a study of a variable difficulty cooperative game for teens with cerebral palsy \cite{hernandez_design_2014} and a program teaching esports to young adults with autism \cite{k_l_nielsen_teaching_2021}. We were surprised at the lack of games ND social computing research in the ACM database, as evidence suggests that ND people tend to be more online than neurotypical people and prefer more structured interaction schemes \cite{ahmed_constrained_2022}. While part of this is because we excluded many simplistic single player therapeutic games, much of the social interaction that occurs across a variety of online multiplayer games (e.g., Minecraft) is consistent with our criteria but was not found in our search. We suspect that the lack of games research with a social computing and CSCW framing is a factor. \textbf{As games continue to expand to massive worlds with individualized experiences, diverse cultures, and complex traditions that are thus far understudied in the context of social computing, games provide an opportunity to study organic online neurodivergent sociality without the pressures of alternatives like networked SNS.} Community behavior, interactions, and moderation within socially open-ended games present an opportunity to develop better understanding of ND users' ideal forms of social governance. However, many of these communities today exist on platforms or games that already embed conventional political defaults and social norms. Future ND social computing work in this space might consider how to offer ND users the agency to redesign alternative, creative re-configurations of social worlds \cite{zhang_policykit_2020}---ones that do not embed conventional or neurotypical norms. 

Ringland’s work in Autcraft finds that pre-existing online communities of ND people develop complex, collaborative moderation schemes with ND and NT peers to create safe and collaborative community spaces. NT adults serve as moderators, but appointed ND “helpers” often serve critical roles in ensuring the smooth operation of the server. While NT adults are again placed in a position of authority over ND speech, these moderators emphasize that they adapt the rules to the ability of the users (e.g., content that would normally constitute spam is allowed so a user with unique needs can use the game's text chat). A neurodiverse moderation team allows for judgments based on preference and ability, bending non-safety-critical rules so users can be expressive \cite{ringland_would_2016}. Digital spaces can provide safe spaces to calm down or vent without judgment and offer variable modes of expression depending on location and social capacity \cite{ringland_place_2019}. \textbf{Game worlds present the opportunity to explore ND sociality in real-time and unconstrained, as these platforms offer a unique blend of non-prescriptive settings with thick context and variable engagement (e.g., lurking, engaging in chat while completing an in-game task, chatting while idle in the world).}

An entire publication is dedicated to ND frustration with NT definitions of being social \cite{ringland_will_2016}. Many users spend hours negotiating complex building projects, discussing interests, and venting about frustrations, yet these social interactions do not seem to “count” in the eyes of NT people in the same way that interacting in person does. In an environment rife with narratives about dwindling attention and screen addiction, the rich relationships that ND people craft online with carefully planned text messages, cooperative projects, and laborious moderation tasks remain invalidated. We conclude our review of games and ND users with the following from Ringland:

\begin{quote}
    “Part of the challenge, particularly as new media continue to emerge, is expanding definitions of sociality that help to weave on- and offline behavior and resonate with the people engaging in them. At one time, talking on the phone was considered an asocial activity compared with face-to-face encounters. As popular opinion changes and more evidence is amassed in the research, \textit{we must be prepared as a scholarly community to understand—if not accept—these community-based definitions of terms we may think we already know.} (emphasis added)” \cite{ringland_will_2016}
\end{quote}

\section{Discussion}
\begin{quote}
    “The cyborg does not dream of community on the model of the organic family, this time without the oedipal project. The cyborg would not recognize the Garden of Eden; it is not made of mud and cannot dream of returning to dust. Perhaps that is why I want to see if cyborgs can subvert the apocalypse of returning to nuclear dust in the manic compulsion to name the Enemy. Cyborgs are not reverent; they do not remember the cosmos. They are wary of holism, but needy for connection—they seem to have a natural feel for united-front politics, but without the vanguard party.” - \textit{A Cyborg Manifesto}, Donna Haraway \cite{haraway_cyborg_1990}
\end{quote}

We now have a depiction of neurodivergent people in existing social computing research. Most often, they are children, and a large majority of these children are male. They are spoken for, not with. They are purely output, expected to tolerate sensors, speak as many words as they can without acknowledgement of their content, and be productive should they appear in corporations or higher education. Even after filtering out hundreds of skills training games and other mainstreaming technologies, the sentiment remains clear---ND social computing research seeks to mainstream or isolate ND people to preserve existing platforms and social spaces. This is present throughout many of these papers, from harmful generalizations about the ``problem'' of disability in their introductions to the celebrations of newfound productivity irrespective of neurodivergent experiences.

A foundational claim of HCI is the commitment to unlocking new dimensions of human relation through the possibilities of computer-facilitated interaction. We live in a period where revolutionary generative artificial intelligence systems challenge what it means to interact with a computer, yet when neurodivergent people interact with these interfaces it is often the hope that they will return to the norms that govern in-person social interaction. Furthermore, there need not be a “discovery” of neurodivergent sociality---neurodivergent
 people have fostered communities online since the modern internet’s inception, from Wrong Planet to Autcraft, yet the labor undertaken to construct these communities is often unrecognized. It should be the goal of future research not to define what neurodivergent sociality should look like, but formalize the invisible labor practices of ND people such that social experience is available to all.

In this section, we draw upon critical disability theory and prior methodological work in HCI to promote accessible research and accessible sociotechnical systems. We begin with reforms for the research process itself and conclude with system-specific recommendations for future work.

\subsection{Access Intimacy and Care Work}
Rather than focusing on the phatic ways we demonstrate care (smiling, maintaining eye-contact, nodding, etc.), disability scholarship outlines several ways in which researchers can genuinely engage with the needs of disabled participants. Deciding to participate in a study can be a monumental task for neurodivergent people; the process of breaking carefully crafted routines, enduring overwhelming sensory experiences, and fulfilling necessary protocol pageantry can be debilitating. Offering open-ended fields for gender options, a range of interview mediums including audio-only and chat interviews, and the choice of whether they would like a trusted person in the room during an interview are all ways to show neurodivergent participants that researchers want them to be comfortable and confident. Mia Mingus terms this exercise \textit{access intimacy}, or “that elusive, hard to describe feeling when someone else ‘gets’ your access needs” \cite{mingus_access_2011}. Especially when working within diagnostic categories like ``autistic teens'' and ``ADHD youth'', it is important to note that the needs of each participant could be radically different. Doing research in this modular way will offer richer answers with more honest criticism, thereby improving the resultant artifacts.

While often critical of the involvement of (implied) neurotypical people in neurodiversity research, we do not consider research with purely neurodivergent participants as intrinsically “better”. \textbf{Our frustration is that current research methods are ignorant of the role of care work, or how disabled people collaboratively meet each others’ access needs.} This may take the form of parent caring for child, as often appears in our corpus, but could also include siblings, teachers, therapists, support staff, and purely online communities that have care practices of their own. Mack et al. explores collaborative, lateral caretaking in the context of chronic illness; while some platforms may be inaccessible due to highly specific sensory needs, communities who have intimate knowledge of these needs can play the role of assistive technology by curating content for their members \cite{mack_chronically_2022}. Much of the work in our corpus positions itself as attempting to understand how neurodivergent people maintain relationships, but many of these relationships are top-down moderation schemes (e.g., parent-child, therapist-patient). While these relationships are certainly worth studying (and do exhibit a form of care work), more attention should be paid toward the collaborative practices of neurodivergent people in care work communities \cite{williams_cyborg_2023}.

Some researchers preemptively attempt to make accommodations for neurodivergent people in research. For example, some researchers may interview participants at their school because it is a familiar environment, while others may offer text-only interviews because there is less pressure to quickly respond. We suggest researchers consider their interview strategy in these terms: Does the medium of the interview significantly impact your ability to ask your specific questions? Would offering different interview mediums to different participants in the same project threaten the validity of the results? How involved are audiovisual cues in your analysis? How might the time dimension of your interview differ based on the format? What is the role of third party assistance in your study, and how might different interview formats facilitate collaborative responses? How might you account for disparities in conversational agency if your participant uses an AAC? Depending on the answers to these questions (and requirements by an institution’s IRB), one may be limited to a specific interview medium. \textbf{Regardless, more detailed explanations of why an interview method was chosen would be helpful for future researchers to understand the pros and cons of a communication medium for a study’s specific neurodivergent population.}

\subsection{Entering and Leaving The Field}
Our corpus is plagued by “parachute” research, as critiqued by Hwang et al. \cite{hwang_exploring_2020}. Researchers often save interactions with ND people for restrictive evaluation phases, giving them little opportunity to give constructive feedback. User involvement is often restricted to a performance task, not allowing for the possibility that, although a user might be able to engage with the device, the experience it provides may not be desirable. This limited involvement of ND people may be because of restrictions on recruiting ND participants for extended periods of time, a desire not to intrude on ND livelihoods, or some combination of the two. Regardless of intention, the result is the same: the technology is designed with specifications that do not necessarily align with the social desires of ND people. Furthermore, after the study, a device designed to be maintained by teachers or staff at a center may be too confusing or laborious, and thus it is abandoned \cite{balasuriya_support_2022}. To make devices that better serve these communities and evaluations that truly assess their use, we ask researchers to more deeply consider how they enter and leave the field.

\textbf{Participants do not spawn at a research site.} Certain language used in recruiting messages and posters may bias toward participants with extremely positive or extremely negative views of their neurodivergent identity. They may have to travel long distances on overwhelming public transit or have to coordinate with others because they may not have a driver's license. Observations may be conducted in rooms flooded with harsh fluorescent lights, scented with overwhelming cleaning supplies, or littered with distracting screens and devices. They may be forced to fill out a form that doesn’t present their gender options, and they may struggle to speak over an accompanying parent who answers questions faster than they can type. After fifteen or twenty minutes, keeping attention may be impossible \cite{mack_anticipate_2022}. Then, months later, they may discover that their words are presented in a completely different context than the one in which they were originally said. Doing care work is a continuous process; it starts long before a study starts and ends long after it has concluded. From recruiting messages to off-script interactions during the research process to descriptions of the neurodivergent identity of interest in the introduction of the paper, the words of the researcher can have negative effects when spoken carelessly and lead to worthwhile collaborations if chosen consciously and respectfully.

\textbf{Participants are more than their words.} Leaving the field is a complex process and is further complicated when partnerships rely on ND participants, parents, and teachers as many of these studies do \cite{scheepmaker_leaving_2021}. If a technology heavily relies on staff or parents to maintain the system, there is a strong possibility that these users will abandon the system if there are no tangible benefits to them. Frequent staff turnover and burnout can make these systems unsustainable long term \cite{balasuriya_support_2022}. Establishing successful relationships might include providing participants a chance to review their quotations and their context or planning long-term projects with opportunities for revision. Designing tools that allow ND people and their collaborators to share the workload of designing and maintaining sociotechnical systems ensures that systems designed for all parties stay working long after researchers have concluded their project, if that is the project's stated goal. If this is not the project's goal, consider what will happen to the constructed prototype or the community that uses it when the current project is completed. The publication itself might not be of use to the participants of the study, but a stable prototype or a commitment to future collaborations may have some use for other researchers who may want to continue this work.

Conferences have a responsibility to provide guidelines or examples for approaching ND social computing. CHI, which has featured more ND social computing papers than any other conference since 2013, only gave one example of such a paper on the CHI 2024 subcommittee page. This paper, SayWAT \cite{boyd_saywat_2016}, has even been criticized by its own first author for its lack of participatory methods with autistic users and the placement of conversational difficulties exclusively on the autistic user without any acknowledgement of their conversation partner \cite{williams_counterventions_2023}. ASSETS supplies a language guide \cite{hanson_writing_2015} that mentions cognitive disability but does not discuss research methods besides referring to ACM policy. CSCW 2024 does not mention marginalization on their “call for papers” page, despite the conference being one of the growing forums for neurodiverse HCI research. Guidance on language, along with examples of participatory ND social computing work, would promote ND social computing research that serves the interests of ND people through meaningful, respectful partnerships.

\subsection{Spectrum-Conscious Social Systems}
A common theme throughout our corpus was a desire to make simple technological solutions that worked for everyone, often at the expense of sophisticated or customized features. This is by no means a new challenge for CSCW; researchers have often struggled to model ambiguous social processes \cite{ackerman_intellectual_2000}. There is a well documented risk to developing systems that may not work for every user, as with sample sizes in the low teens, one user’s struggles with a system could be targeted by reviewers and negatively impact a researcher’s chance of publication \cite{bernstein_trouble_2011}. As noted, research in our corpus has such small sample sizes that a single failure for a particular user could be devastating for a project. However, in the context of autism, ADHD, and several other ND diagnostic categories, the aim of designing for everyone within a diagnostic category is impractical and misguided. As discussed previously, the needs of nonverbal users are often unmet in order to design for and test systems with more privileged neurodivergent participants in technology adjacent roles (e.g., software engineers, college students, online activists). \textbf{A purely medical model of disability often neglects the needs of ND people in these spaces, whereas a purely social model neglects the needs of those with alternative social needs who need significant assistance to complete daily tasks.} ND social computing research has to use both and do so appropriately.

Motivated by neurodiversity’s embrace of spectrums of ability, we ask system designers to consider novel ways for users to modify and coordinate their system to fit unique needs. As addressed by research in chronic illness and HCI \cite{mack_chronically_2022}, chronically ill and disabled people often have  unique experiences that cannot be addressed individually at a system level. Additionally, these experiences can change depending on their aptitude to perform the work of curation on a particular day, necessitating systems that adapt quickly and often \cite{das_towards_2021}. While one autistic adult may be engrossed in food culture and spend hours on forums discussing recipes, another autistic adult may be sickened by food pictures to the degree that they spend hours curating block-lists to remove them from their feed. While an ADHD person may enjoy being absorbed in an endless feed on a Saturday afternoon, this may be a source of inescapable anxiety on a Monday night before a deliverable must be submitted. In advocating for features to address this barrier to being social online, we draw upon critical race theory and the concept of interest convergence, which warns against providing rights or features only if these additions benefit everyone \cite{bell_brown_1980}. Rather than platform-wide restrictions on food, which would be undesirable and infeasible, systems should feature collaborative and transparent settings that allow ND users and their communities to curate palatable content~\cite{jhaver_designing_2022}, even in ways that might seem undesirable. \textbf{These interdependent systems must also provide structured methods to ask for assistance while assuring the privacy of users that rely on others to curate feeds,  help respond to harassment, and get work done.}

Consider also how these features might require negotiation between friends building collaborative filters, parents or caretakers working with children to build protective measures, or some combination of these two scenarios. ND people may want the help of a parent, sibling, or friend on a particularly difficult day and ask for this assistance explicitly, or they may desire this assistance and be too overwhelmed to phrase this ask. On another day, they may wish to be left alone entirely. Future work should explore how both sides of this relationship negotiate their involvement in each others' social experiences as they make these decisions concerning privacy and safety.

\subsection{Double Empathy in Trust and Safety}
As discussed previously, most of the implied relationships in ND social computing systems are between parents and children or doctors and patients. There are some instances in which these systems can be absolutely critical, such as interventions for challenging behaviors and self harm \cite{jo_geniauti_2022}. There are others, however, which track location and biometric data to a degree that is invasive and inappropriate for ND young adults and adults, irrespective of their environment. By including more ND people in the requirements and design phases of these technologies, researchers can better encode boundaries that ND users can collaboratively set to serve their interests without breaching their privacy. Different methods will be necessary for different populations; for example, ADHD adults may prefer an in-person workshop to prototype a system from the ground-up across several sessions, whereas several phases of iterative prototyping may be helpful for getting feedback from teens with  intellectual and developmental disabilities (IDD) \cite{xu_rove_2014}. Both studies could provide opportunities to redesign social systems as collaborators rather than evaluators. Some areas of focus identified by the analysis of our corpus include more transparent and fine-grained privacy settings for SNS platforms, moderation strategies that allow ND users to learn without being permanently blocked, and further investigation into social sensing systems for both NT and ND users that are truly non-prescriptive \cite{fiesler_creativity_2019, barros_pena_my_2023, ringland_would_2016}.

More research into ND/NT and ND/ND friendships and collaborative strategies could be a promising direction for the design of ``friendsourcing'' systems on social platforms. Games research, although presently limited by an overwhelming number of skill training games \cite{spiel_purpose_2021}, is a promising domain to analyze these dynamics. Online gaming is an excellent blend of thick platform context, high customization, and variable degrees of engagement, and many ND people find games with others to be a highly rewarding form of social interaction. After conducting significant care work, they could be the best opportunity to, in the words of Ringland, “understand—if not accept—these community-based definitions” of being social \cite{ringland_will_2016}.

Double empathy asks us to think more deeply about both sides of networked interactions online. ND social computing users often have complex invisible labor practices that make online social experience palatable. \textbf{Ensuring the recognition of these practices through online support and features by targeting the recipients of ND social output online is often understudied in favor of client-side solutions for ND users.} We ask researchers to consider both sides of these networked interactions and assert that only platform-wide solutions will create doubly-empathetic technology.

\section{Conclusion}
This paper presents the first comprehensive review of social computing with and for neurodivergent people. Through our analysis of 84 publications from the past ten years, we confirmed that ND social computing research focuses on the study of children, often relies on female caretakers of male children (thereby perpetuating gendered labor practices and limiting expression outside the gender binary), often involves ND people exclusively in the evaluation of the artifact, and is largely not inclusive of communication mediums beyond verbal speech. Requirements for ``being social'' are often devised by parents or contextualized within institutions like schools, workplaces, and skill training centers, thus adopting a definition for sociality that is inconsistent with ND sociality in that it is restrictive and seeks to mainstream ND user behavior. NT people often infringe on the conversational agency of ND people in ND social computing systems and in the research process, and there is a lack of attention toward the effort required by NT users of these systems that maintain these infrastructures in addition to participating in interactions. 

Using the framing of double empathy, we envisioned how technology has the capability to redefine normative social experience and has several features that may be desirable to ND people. In doing so, we identified possible areas of future research, inclusive participatory research methods to collaborate with ND people, more targeted contributions given the high degrees of variance within diagnostic categories of neurodiversity, and explainable trust and safety for ND people and their communities. In this way, we call not for a revolution but for a reaffirmation of the values of the HCI community. The promise of going beyond conventions should naturally extend to the neurodivergent people whom these conventions harm, and for this reason we have an obligation to collaborate with neurodiverse communities in building sociotechnical systems that go beyond our accepted visions of ``being social''.

\bibliographystyle{ACM-Reference-Format}
\bibliography{ndsc}

\section{Appendix}
\subsection{Query}
Query is based on a version of Spiel and Gerling's query in their review of neurodiverse games \cite{spiel_purpose_2021}, with some omissions based on whether a corresponding neurodivergent identity appeared in the search and additions to represent identities that appeared in independent searches for relevant neurodivergent identities. Neurodiverse identities were included with a focus on including experiences that modify normative conceptions of enjoyable social experience. See Table 3 for the full query.
\begin{table}[ht]
\small
    \centering
    \begin{tabular}{{|p{0.35\textwidth}|p{0.1\textwidth}|p{0.4\textwidth}|}}
    \hline
    Neurodiversity Term & AND & Social Computing term \\ \hline
    autism OR autistic OR autistics OR “Autism Spectrum Disorder” OR ASD OR "attention deficit disorder" OR "attention deficit hyperactivity disorder" OR “cognitive disability” OR “cognitive disabilities” OR “cognitive impairment” OR “cognitive impairments” OR “learning disability” OR “learning disabilities” OR “learning impairment” OR “learning impairments” OR "intellectual disability" OR “intellectual disabilities” OR “intellectual impairment” OR “intellectual impairments” OR “special needs” OR “developmental disability” OR “developmental disabilities” OR “developmental impairment” OR “developmental impairments” OR “intellectual and developmental disability” OR IDD OR “complex needs” OR “complex disability” OR “complex disabilities” OR “complex impairment” OR “complex impairments” OR ADHD OR dyslexia OR dyscalculia OR dysgraphia OR "trisomy 21" OR "down syndrome" OR Asperger OR “cerebral palsy” OR “obsessive compulsive disorder” OR “OCD” OR neurodiversity OR neurodivergent OR neurodiverse & AND & "social computing" OR "SNS" OR "social media" OR social OR society OR online OR collaboration OR communication \\ \hline
    \end{tabular}
    \caption{Query used to search the ACM database. If any of these terms appeared in the title, abstract or author keywords for the article, it was included in the intial collection of 863 search results. Note also that inclusion of any of these terms in the search does not mean the authors condone their usage. While there is much debate about what constitutes neurodivergent identity, the authors purposely adopted as broad a definition as possible in order to most accurately represent the design space for social needs that differ from the norm.}
    \label{query}
\end{table}
\newpage
\subsection{Codebook}
The following categories and options were used to classify each paper.
\subsubsection{Year}
Note the year of publication, as listed on the article.
\subsubsection{Continent/Country of Institution(s)}
Select one of the following: Africa, Asia, Australia, Europe, North America, South America. \newline
    Then, list the countries in which all institutions are located in parenthesis. \newline
    Ex. North America (United States, Canada), Europe (UK)
\subsubsection{Contribution Type (Wobbrock 2015) \cite{wobbrock_seven_2015}}  Select one of the following contribution types:
    \begin{description}
        \item[\textbf{Empirical}] New findings based on systematically gathered data. Empirical contributions may be quantitative or qualitative (or mixed), and usually follow from scientific studies of various kinds (e.g., laboratory, field, ethnographic, etc.)
        \item[\textbf{Artifact}] Novel systems, architectures, tools, techniques, or designs that reveal new opportunities, enable new outcomes, facilitate new insights or explorations, or impel us to consider new possible futures
        \item[\textbf{Mixed}] An empirical, need-finding study is conducted before constructing an artifact
    \end{description}
\subsubsection{System Studied} Select one of the following research areas:
    \begin{description}
        \item[\textbf{Augmentative and Alternative Communication:}] Paper focuses on alternative forms of communication, specifically for neurodivergent users. These devices directly emit the speech of ND people.
        \item[\textbf{Digital Collaboration:}] Platforms for conversation with defined partners, like text and video chat. Some examples would be video conferencing, SMS messaging, VR, or systems that allow people to share biometric information.
        \item[\textbf{Games:}] Paper focuses on games where users interact primarily in the context of the game (ex. Minecraft)
        \item[\textbf{Sensors for Social Feedback:}] Paper focuses on systems that display latent social information in an alternative way. These include emotion recognition and distance detection systems
        \item[\textbf{Social Network Services:}] Paper focuses on experiences with or a novel social network service (SNS), as defined by Oxford: “use of Internet technologies to link together users of the Internet who have common interests”. A platform which allows people to chat, comment, and share media is probably an SNS.
        \item[\textbf{Workplace Culture and Organizations:}] System of interest is a program or organization that uses social computing technology and is focused on the needs of ND people. These include summer camp programs and workplace studies.
    \end{description}
\subsubsection{Neurodiverse Focus} List the neurodivergent identities of all participants in the study and any collective categories that are also mentioned. \newline
\begin{description}
        \item[\textbf{Neurodiverse:}] Paper focuses on a range of several neurodivergent identities.
        \item[\textbf{ADHD:}] Attention Deficit Hyperactivity Disorder, characterized by hyperactivity, impulsivity and inattention. The ``inattentive'' form of ADHD has often been neglected because of ADHD's reputation as a childhood disorder, but it is important to note that ADHD in adults and women may present itself differently from stereotypes of ADHD in boys \cite{spiel_adhd_2022}.
        \item[\textbf{Autism:}] A neurodiverse identity commonly characterized by preference for routine and interest-based social experiences. Autistic people often have sensory preferences that make bright lights, loud spaces, and face-to-face conversation undesirable \cite{ochs_autistic_2010, das_towards_2021}.
        \item[\textbf{Asperger's Syndrome:}] A previously used diagnostic category now recognized to be another expression of autism \cite{morris_understanding_2015}.
        \item[\textbf{Cerebral Palsy:}] A group of disorders that affects the development of motor function, often requiring communication aides or adaptive controls to use technologies  \cite{hernandez_design_2014}.
        \item[Cognitive Disability or Intellectual Disability]: A collection of neurodiverse identities related to cognitive processes like decision making \cite{piper_technological_2016}.
        \item[Learning Disability]: A collection of neurodiverse identities related to reading and verbal communication \cite{hooper_building_2015}.
        \item[Dyslexia]: Historically viewed as a difficulty with learning by reading text, dyslexia is increasingly considered as a preference for alternate modes of learning and processing beyond written words  \cite{vezzoli_dyslexia_2017}.
\end{description}
\subsubsection{Research Perspective} Choose the statement that best represents the perspective researchers take toward neurodiversity, neurodivergent people, and the role of their social computing system in their lives.
    \begin{description}
        \item[\textbf{Symptoms:}] technology is targeted at mitigating or minimizing an expression of a particular symptom attributed to the disability, independent of a specific environment or context; description focuses on deficit to be compensated (ex. Minimize fidgeting, enforce eye contact, intervention)
        \item[\textbf{Understanding:}] stated goal of the project is to understand the practices of neurodivergent people in existing spaces
        \item[\textbf{Productivity:}] analysis or technology is situated in a specific environment (work, school, etc) and/or its stated goal is to improve function within that given setting (ex. “Improve focus in school”, “limit distractions at work”, etc)
        \item[\textbf{Expression:}] technology is meant to enhance or augment expression of feelings or needs in a non prescriptive way
        \item[\textbf{Relationships:}] technology is designed to compliment interactions between neurodivergent people and their families, friends, therapists, etc. Technology or study is meant to help ND people feel safe and included on social platforms.
    \end{description}
\subsubsection{Positionality} \begin{description}
        \item[\textbf{None:}] No positionality statement is included. There is no reference to the author’s identity within the paper itself. This code should be applied even if the author publicly discloses neurodivergence in other contexts (ex. personal social media, conferences, etc)
        \item[\textbf{Same:}] An author states that their neurodivergent identity matches the identity studied. \newline
        Ex: "The lead researcher—who is autistic, has degrees in psychology and human–computer interaction, and experience conducting research with neurodivergent people—conducted a thematic analysis on autistic and non-autistic participants’ transcripts separately" \cite{rocheleau_privacy_2022}
        \item[\textbf{Different:}] An author discloses specific identity characteristics with respect to neurodivergence but they differ from the population of interest. \newline
        Ex: “At least one author is neurodivergent; none of the authors experience dyslexia” \cite{wang_invisible_2022}
    \end{description}
\subsubsection{Study Description} Complete this category for each listed sample in the study, and separately for each neurodiverse (ND) and neurotypical (NT) subgroup. If any category is not applicable, fill in N/A. \newline
    If there are multiple applicable answers, list all that apply. Note that there may be multiple samples within each paper.
    \begin{description}
        \item[\textbf{Sample Size:}] Number of participants in the final analysis (be sure to check if any participants were excluded before the culmination of the study) OR N/A for ethnography, indeterminate number, scraping of public data
        \item[\textbf{Identifiers:}] If any of the following information is presented for every participant, please use the appropriate abbreviation in a comma-separated list:
        \begin{itemize}
            \item Gender (Binary): GB
            \item Gender (Nonbinary): GNB
            \item Gendered Pseudonym: GP
            \item Race: R
            \item Age: A
            \item IQ: IQ
            \item High-Functioning/Low-Functioning: HF/LF
            \item Employment/Education: E
            \item Nationality: N
            \item Familial Relationship: F
            \item None: None
        \end{itemize}
        \item[\textbf{Age:}] Age range for the participants in the study formatted as [MIN AGE]/[MAX AGE] OR one of the following categories should they be explicitly mentioned or implied by other identity characteristics:
        \begin{description}
            \item[\textbf{Children:}] Target demographic is stated to be “children” or study includes those aged 0-12
            \item[\textbf{Adolescent:}] Target demographic is stated to be “adolescents”, “teenagers”, or “young adults” or study includes those aged 13-20
            \item[\textbf{Adults:}] Target demographic is stated to be “adults” or those 21-49 years old
            \item[\textbf{Older Adults:}] Target demographic is stated to be, “older adults”, or adults older than 50
            \item[\textbf{None:}] There is no target age mentioned
        \end{description}
        \item[\textbf{Gender Distribution:}] If gender is not noted, code as N/A. Otherwise, select one of the following.
        \begin{description}
            \item[\textbf{Male Majority:}] Over 50\% of participants are male
            \item[\textbf{Female Majority:}] Over 50\% of participants are female
            \item[\textbf{No Majority:}] There is no outright majority
            \item[\textbf{Balanced:}] There is an even distribution of male and female participants
        \end{description}
        \item[\textbf{Diagnosis: }] Select one of the following options:
        \begin{description}
            \item[\textbf{Yes:}] Formal diagnosis via a medical professional or formal diagnostic criteria is required, OR researchers select participants based on the presentation of a set of symptoms that form a diagnostic criteria, OR participants are recruited from a location that requires diagnosis
            \item[\textbf{No:}] Self identification is explicitly allowed
            \item[\textbf{NT:}] Assumed to be neurotypical
            \item[\textbf{Unclear:}] Authors do not clarify whether diagnosis was or was not required
        \end{description}
        \item[\textbf{Sourcing: }] Select one of the following:
        \begin{description}
            \item[\textbf{Decentralized:}] Authors recruited participants from personal networks, using techniques like snowball sampling, or general internet channels outside the context of neurodiversity. Recruitment methods that use a blend of physical and online spaces are considered decentralized. \newline
            Ex: "We invited parents of autistic as well as non-autistic children via social media, including Facebook and Twitter. In addition, parents of autistic children were approached by contacts with relevant organizations..." \cite{macmillan_are_2020}
            \item[\textbf{Centralized (Physical):}] Authors only recruited from residents or members of a physical center or set of similar physical centers (ex. school, assisted living facility, charity organization, etc) \newline
            Ex: "16 children and 12 teachers were recruited from a small private school that specializes in educating children with autism and other communication social difficulties using the Floortime Model" \cite{mokashi_exploration_2013}
            \item[\textbf{Centralized (Online):}] Authors recruited from a space marked specifically for neurodivergent people (ex. “r/Autism”, “r/ADHD”) \newline
            Ex: “This paper reports on results from an on-going, qualitative digital study of an online community that has grown around a Minecraft server known as Autcraft.” \cite{ringland_would_2016}
Unclear: Recruitment process is vague or references a recruitment service without elaborating on methods
        \end{description}
        \item[\textbf{Consent: }] Select one of the following:
        \begin{description}
            \item[\textbf{Full:}] Participants are given the choice of whether or not to participate in absence of significant influence from trusted community members. It is stated or implied that participants have control over whether others are in the room during studies and interviews. \newline
            Ex: Participants were contacted and independently scheduled interviews. Parents signed a consent form only once they were at the physical location and participants were given the option to conduct the interview with a parent in the room or on their own. \cite{rocheleau_privacy_2022}
            \item[\textbf{Proxy:}] Parents, therapists, or other trusted figures are contacted about and facilitate participation BEFORE (and sometimes instead of) individuals themselves. This will be the case of most studies involving children, as well as those conducted with centralized geographic sourcing (since the institutional affiliation influences participation). \newline
            Ex: " The researchers will ask for consent from the teachers and the parents/guardians of the children if they will allow the students to partake in the initial testing" \cite{samonte_tap--talk_2020}
            \item[\textbf{None:}] Study scrapes data from publicly available message boards or accounts. Participants are not aware that their contributions are being used in a study. \newline
            Ex: "We investigated a large, popular online autism forum with more than 6,000 registered members and 19 public discussion boards. We collected data only from 15 publicly-accessible discussion boards in this forum" \cite{hong_towards_2015}
            \item[\textbf{Unclear:}] Recruitment process is vague or references a recruitment service without elaborating on methods
        \end{description}
        \item[\textbf{Method:}] Select one of the following research methods and communication schemes for each phase of the study:
        \begin{description}
            \item[\textbf{Observation:}] ND participants are watched as they use a technology. There is no prescriptive or ulterior motive designated for the use of said technology.
            \item[\textbf{Task Performance:}] ND participants are expected to perform a technology mediated task
            \item[\textbf{(In Person/Chat/Teleconference)}] Interview: primary focus of analysis is in the format of a synchronous Q\&A session
            \item[\textbf{Ethnography:}] a holistic account of an ND community
            \item[\textbf{Questionnaire:}] a series of questions meant to assess the needs of ND people directly
            \item[\textbf{Proxy Questionnaire:}] a series of questions meant to assess the needs of ND people indirectly
            \item[\textbf{Workshop:}] a group of ND people design a technology together
            \item[\textbf{Proxy Workshop:}] a group of parents, caretakers, and/or experts design a technology for ND people together
            \item[\textbf{Other}]
        \end{description}
    \end{description}
\subsubsection{Evaluation Criteria} Select both an evaluation criteria and the perspective of the evaluation (ex. Satisfaction by ND):
    \begin{description}
        \item[\textbf{Comparison (Researcher/Proxy):}] Evaluation is based on a comparison between NT and ND participants by proxies or researchers
        \item[\textbf{Performance (Researcher/Proxy):}] Evaluation is qualitative, but usefulness is assessed by someone other than the ND user
        \item[\textbf{Satisfaction (ND/Proxy):}] Evaluation mainly focuses on whether users and stakeholders enjoy using the technology
    \end{description}
\subsubsection{Participation Level (Çorlu et al. 2017) \cite{corlu_involving_2017}} Note all the phases of research in which ND people participated (ex. RE):
    \begin{description}
        \item[\textbf{R:}] ND people participate in the development of requirements for the technology
        \item[\textbf{D:}] ND people participate throughout the design process, either as members of the team or stakeholders that give feedback on prototypes
        \item[\textbf{E:}] ND people are present for the evaluation of the technology
    \end{description}
\subsubsection{Future Work \& Limitations}
Note the mention of any of the following topics in the future work and limitations sections:
    \begin{description}
        \item[\textbf{Continuation (Satisfied):}] The authors articulate a desire to continue studying the domain of the publication of interest (ex. “Conduct more user tests of XYZ”, “change the design of XYZ and test with ABC”). There is a future publication that accomplishes this task.
        \item[\textbf{Continuation (Unsatisfied):}] The authors articulate a desire to continue studying the domain of the publication of interest (ex. “Conduct more user tests of XYZ”, “change the design of XYZ and test with ABC”). There is no future publication that accomplishes this task.
        \item[\textbf{Co-Design:}] The future work section mentions a desire for more co-design with neurodivergent participants.
        \item[\textbf{Generalizability:}] Authors express an interest to make their work more widely applicable OR say their work is limited by small/specific sample
        \item[\textbf{Collaboration:}] Authors express desire to promote collaboration with neurodivergent researchers
        \item[\textbf{None:}] There is no mention of future work
    \end{description}

\section{Acknowledgements}
The authors would like to thank the members of the Social Futures Lab for their feedback throughout this process. 
We would also like to thank Jack Ruzekowicz for his constructive feedback and insight.
Funding for this project was provided by the National Science Foundation via the 2023 DUB REU Program at the University of Washington.

\end{document}